\documentclass[reprint,amsmath,amssymb,aps,prx,superscriptaddress, showpacs, eprint]{revtex4-2}
\usepackage[utf8]{inputenc}

\usepackage{amsmath,amssymb,ascmac,fancybox, multirow}
\usepackage{color}
\usepackage{graphicx}
\usepackage{url}
\usepackage{siunitx}
\usepackage{bm}
\usepackage{physics}
\usepackage{mathtools}
\usepackage{hyperref}
\usepackage[T1]{fontenc}
\usepackage{ulem}
\graphicspath{figures/}

\newcommand{\ani}[1]{{#1}}
\newcommand{\cre}[1]{{#1}^\dagger}

\renewcommand{\vec}{\vb*}

\newcommand{\red}[1]{#1}

\usepackage{makecell}

\begin{document}
\date{\today}
\title{Quantum weight: A fundamental property of quantum many-body systems}
\author{Yugo Onishi}
\affiliation{Department of Physics, Massachusetts Institute of Technology, Cambridge, MA 02139, USA}
\author{Liang Fu}
\affiliation{Department of Physics, Massachusetts Institute of Technology, Cambridge, MA 02139, USA}

\begin{abstract}
We introduce the concept of quantum weight as a ground state property of quantum many-body systems that is encoded in the static structure factor and characterizes density fluctuation at long wavelengths. The quantum weight carries a wealth of information about dielectric responses and optical properties of the system, and is closely related to its quantum geometry. For systems with short-range interactions or low-dimensional Coulomb systems, we show that the many-body quantum metric (which measures the change of the ground state under twisted boundary conditions) can be determined directly from the quantum weight. \red{Notably, the quantum weight is a property of a single ground state and independent of boundary conditions in the thermodynamic limit. Our finding thus enables direct experimental measurement and numerical calculation of many-body quantum metric.}
On the other hand, for three-dimensional Coulomb systems, we \red{show that the quantum weight is distinct from the many-body quantum metric due to dielectric screening in three dimensions.} We further 
use dielectric sum rules to derive upper and lower bounds on their quantum weight in terms of electron density, static dielectric constant, and plasmon energy.   
Our work highlights quantum weight as a fundamental material parameter, which can be experimentally determined by x-ray scattering or electron loss spectroscopy. 
\end{abstract}

\maketitle

\section{Introduction}
The ground state wavefunction of quantum many-body systems encodes a wealth of information about their thermodynamic and transport properties. 
Ground state correlation functions can reveal the existence of long-range order, such as magnetism, superconductivity, and charge density waves. 
The topology of the ground state distinguishes topologically distinct phases of insulators and determines 
the quantization of physical properties such as the Hall conductivity~\cite{Thouless1982, niu_quantized_1985}. 

In this work, we introduce a ground state property of quantum many-body systems that governs density fluctuation at long wavelengths, which we call quantum weight. The quantum weight is defined with the ground-state static (or equal-time) structure factor $S_{\vec{q}}$ at small wavevectors $q\rightarrow 0$, and characterizes quantum fluctuations in electron density. In the classical limit $\hbar\rightarrow 0$, the static structure factor of periodic electron systems exhibits Bragg peaks at reciprocal lattice vectors and vanishes everywhere else. In contrast, in quantum many-body systems, quantum fluctuations in the ground state allow $S_{\vec{q}}$ to be finite away from the reciprocal lattice vectors\red{, giving rise to a finite quantum weight}. Thus, the quantum weight quantifies the degree of ``quantumness'' of the system \red{on macroscopic length scales}.

As part of the static structure factor, the quantum weight can be directly measured through X-ray scattering or electron loss spectroscopy, and calculated numerically from the ground state wavefunction alone. We will show that, despite its simplicity, the quantum weight contains a wealth of information about thermodynamic, dynamical, quantum geometric, and topological properties of the system.

\begin{figure}
    \centering
    \includegraphics[width=1.0\columnwidth]{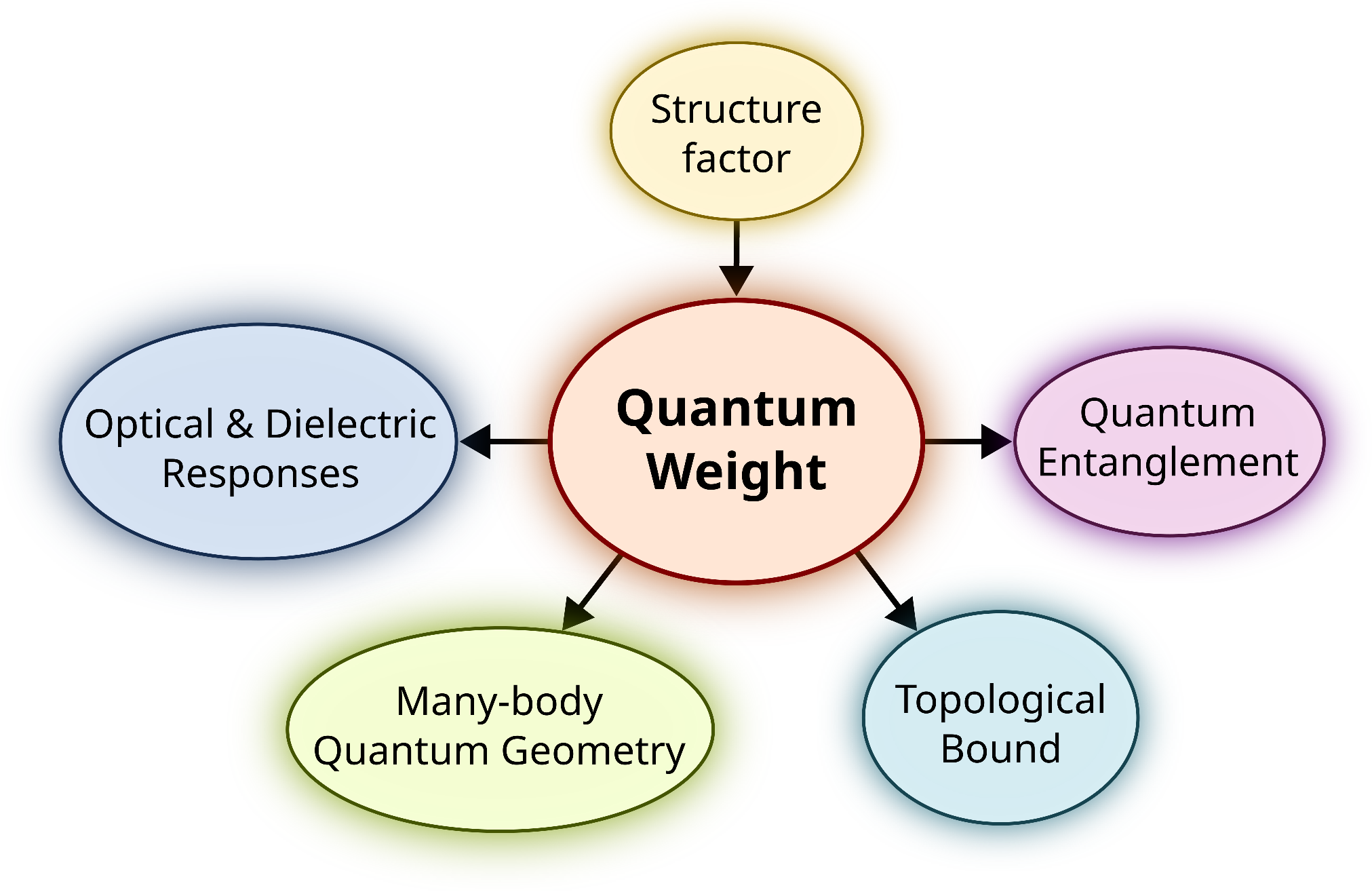}
    \caption{Relation between quantum weight and other properties of quantum many-body systems.}
    \label{fig:quantum_weight}
\end{figure}

The quantum weight is closely related to dielectric responses. Using sum rules for dynamical density response, we derive {\it rigorous} upper and lower bounds on the quantum weight of three-dimensional (3D) Coulomb systems based on the plasmon energy, static dielectric constant, and electron density. We show that the upper and lower bounds are remarkably close to each other for various semiconductors and insulators, thus providing accurate estimates of their quantum weights.

For systems with short-range interaction or Coulomb systems in one or two dimensions, the long-wavelength density response is directly related to the optical conductivity $\sigma(\vec{q}, \omega)$ at $\vec{q}\to 0$ through the continuity relation. Using this relation, we derive a sum rule that relates the quantum weight to optical conductivity.
Further combining this result with the Souza-Wilkins-Martin (SWM) sum rule~\cite{souza_polarization_2000}, we establish a universal relation between the quantum weight and the many-body quantum metric that describes the response of the ground state wavefunction to twisted boundary condition. 
Our finding shows for the first time that the quantum geometry of general quantum many-body systems can be directly determined from the equal-time ground-state correlation function. \red{This connection provides a computational shortcut for obtaining the many-body quantum metric from a single ground state, which is independent of boundary conditions in the thermodynamic limit. Moreover, since optical conductivity and structure factor are easily measurable, this relation offers a direct way to measure many-body quantum geometry of two-dimensional materials. }

The connection with the quantum metric allows us to further relate the quantum weight to the topology of the ground state. As shown in our recent work~\cite{onishi_topological_2024}, the quantum weight has a lower bound determined solely by the ground-state Chern number. Here we show that this bound can be understood as a geometric inequality between the many-body quantum metric and the many-body Berry curvature.

Our work establishes quantum weight as a central quantity that connects various physical observables of quantum many-body systems (Fig.~\ref{fig:quantum_weight}) and demonstrates that it is a powerful tool for studying the electronic structure and dynamical responses of solids.

This work is organized as follows. In Sec.~\ref{sec:Sq_band_geometry}, we define quantum weight in terms of static structure factor and show its relation to band geometry in band insulators. In Sec.~\ref{sec:noint_sumrule}, we discuss a sum rule relating optical responses to quantum weight in systems with short-range interaction. Using this sum rule, we establish the relation between the quantum weight and many-body quantum geometry. In Sec.~\ref{sec:topological_bound}, we discuss a recently shown topological bound on quantum weight~\cite{onishi_topological_2024} in terms of many-body quantum geometry. Sec.~\ref{sec:wigner_solids} and Sec.~\ref{sec:Quantum_weight_Coulomb} focus on quantum weight in Coulomb systems. We first provide an illustrative example of 3D Coulomb systems in Sec.~\ref{sec:wigner_solids}, and then present a general theory of Coulomb systems in Sec.~\ref{sec:Quantum_weight_Coulomb}. In Sec.~\ref{sec:bound}, we derive exact bounds on quantum weight and apply them to real materials. Sec.~\ref{sec:entanglement} concludes with a future outlook. 

\begin{widetext}
\section{Structure factor, Quantum weight, and Band geometry} \label{sec:Sq_band_geometry}
\end{widetext}

In order to quantify the amount of position fluctuation in a many-particle system at zero temperature, we consider the static structure factor which measures equal-time density-density correlation in the ground state: 
$$S_{\vec{q}} \equiv \frac{1}{V} \expval{\hat{n}_{\vec{q}} \hat{n}_{-\vec{q}}}.$$ 
Here, $\hat{n}_{\vec{q}}=\int\dd{\vec{r}}e^{-i\vec{q}\vdot\vec{r}}\hat{n}(\vec{r})$ is the number density operator with wavevector $\vec{q}$ and $V$ is the volume of the system.  For continuous systems (with or without periodic potential), 
$\vec q$ denotes the real wavevector in the range $[-\infty, \infty]$ (rather than quasimomentum in reduced zone), 
and $\hat{n}_{\vec{q}}=\sum_{\vec{k}} c_{\vec{k}+\vec{q}}^\dagger c_{\vec{k}}$ where $c_{\vec{k}}$ is associated with a {\it plane wave} mode (rather than a Bloch state). \red{Note that $S_{\vec{q}}$ thus defined (``full structure factor'') is {\it not} periodic in $\vec q$ even for periodic solids.} For lattice models, $\hat{n}_{\vec{q}}=\sum_i e^{i\vec{q}\vdot\vec{r}_i}\hat{n}_i$ can be defined with the density operator $\hat{n}_i$ on site $i$, in which case only $\vec q$ modulo reciprocal lattice vector is meaningful.  

For a periodic system, the static structure factor exhibits Bragg peaks at reciprocal lattice vectors denoted as $\vec{G}$, where $S_{\vec{G}}$ is proportional to the volume of the system. In addition, the quantum uncertainty in particle position allows the structure factor at other wavevectors $S_{\vec{q} \neq \vec{G}}$ to be nonzero and of order $\order{1}$. 
We shall focus on the leading order behavior of $S_{\vec q}$ at small finite $\vec q$, which describes long-wavelength quantum fluctuations in the ground state.

This work considers quantum many-body systems in which $S_{\vec q}$ is proportional to $q^2$ at small wavevector:
\begin{align}
    S_{\vec{q}} 
    &= \frac{1}{2\pi} K_{\alpha\beta}q_{\alpha}q_{\beta} + \dots, \label{eq:QW_def}
\end{align}
where $\alpha, \beta$ are spatial indices.
 As we shall show later, the static structure factor $S(q)$ 
of insulators and of 3D metals with long-range Coulomb interaction is generally proportional to $q^2$ at small $q$.    
We call the quadratic coefficient $K$ defined by Eq.~\eqref{eq:QW_def} \textit{quantum weight}.   
$K$ has the unit of length to the power of $2-d$, where $d$ is the spatial dimension of the system. In particular, the quantum weight is dimensionless in two dimensions.

To gain intuition about the quantum weight, let us start with an example of noninteracting electrons in a deep periodic potential. The classical ground state at $\hbar=0$ is simply a lattice of electron point particles located at the periodic potential minima. Once electron's kinetic energy is taken into account, there will be quantum fluctuation in electron's position due to the Heisenberg uncertainty principle, which makes the quantum ground state different from the classical one. 

Assuming there is one (spinless) electron per unit cell, the electrons are localized at the periodic potential minima. Expanding the potential around each minima $\vec{R}$ as $V(\vec{r}+\vec{R})=m\omega_0^2 r^2/(2m)$, we can effectively describe the system as an array of harmonic oscillators. The ground state of this system is then given by the Slater determinant of the $n=0$ wavefunctions for each oscillator at $\vec{R}$: $\psi(\vec{r}) =  \qty(m\omega_0/(\pi\hbar))^{d/4} \exp(-m\omega_0 \abs{\vec{r}-\vec{R}}^2/(2\hbar))$. In this case, the structure factor is given by 
\begin{align}
    S_{\vec{q}} = \frac{1}{V_{UC}}(1-e^{-\hbar q^2/(2m\omega_0)}) = \frac{\hbar q^2}{2m\omega_0 V_{UC}} + \dots, \label{eq:HO_Sq}
\end{align}
and thus the quantum weight is
\begin{align}
    K_{\alpha\beta} &= \delta_{\alpha\beta} \frac{\pi \hbar}{m\omega_0 V_{UC}} = 
    \delta_{\alpha\beta} \frac{2\pi \langle x^2 \rangle }{V_{UC}}
\end{align}
with $V_{UC}$ the unit cell volume and $\expval{x^2} = \hbar/(2m\omega_0)$ the position fluctuation of one electron in a harmonic oscillator. In this example, the quantum weight is proportional to the ratio between the intra-unit-cell position fluctuation and the unit cell volume.   

More generally, we can calculate the structure factor and the quantum weight for noninteracting band insulators. It is straightforward to show that the static structure factor for finite $\vec{q}$ is given by (see Supplemental Materials for details)
\begin{align}
    S_{\vec{q}} &= \int_{\rm BZ}\frac{\dd^d{\vec{k}}}{(2\pi)^d} \Tr[P(\vec{k})(P(\vec{k})-P(\vec{k}+\vec{q}))], \label{eq:BI_Sq}
\end{align}
where $P(\vec{k})=\sum_{n}^{\rm occ} \ketbra{u_{n\vec{k}}}{u_{n\vec{k}}}$ is the projection operator onto the occupied bands at wavevector $\vec{k}$ with $\ket{u_{n\vec{k}}}$ the cell-periodic Bloch wavefunction for $n$-th band. Notably, the static structure factor is determined only by the Bloch wavefunction $\ket{u_{n\vec{k}}}$ of the occupied bands and does not depend on the energy dispersion or excited states at all. In particular, the long-wavelength behavior of the structure factor is captured by the band geometry. The band geometry of the occupied bands is characterized by the quantum geometric tensor defined as~\cite{provost_riemannian_1980}
\begin{align}
    Q_{\alpha\beta}(\vec{k}) &\equiv \Tr[(\partial_{\alpha}P)(1-P)(\partial_{\beta}P)].
\end{align}
Its real part $g(\vec{k})=\Re Q(\vec{k})$ defines the quantum metric, while its imaginary part $\Omega(\vec{k})=-2\Im Q(\vec{k})$ defines the Berry curvature of the occupied bands. Noting that $\Tr[P\partial_{\alpha}P]=0$ which follows from $P^2=P$ and $\Tr P=\mathrm{const.}$, the quantum weight defined by the $q^2$ term in $S_{\vec q}$ can be written as the integrated quantum metric: 
\begin{align}
    K_{\mu\nu} &= 2\pi\int_{\rm BZ}\frac{\dd^d{\vec{k}}}{(2\pi)^d} g_{\mu\nu}(\vec{k}), \label{K-BI}
\end{align}
where we used a relation $g_{\mu\nu}(\vec{k})=(1/2)\Tr[(\partial_{\mu}P(\vec{k}))(\partial_{\nu}P(\vec{k}))]$. The integrated quantum metric was originally introduced in the study of Wannier functions and gives the gauge-invariant part of their spread \cite{marzari_maximally_1997}. Thus, the quantum weight defined by the static structure factor is a natural generalization of the position fluctuation to extended systems.

As we have shown, the quantum weight quantifies the quantum fluctuation of the electron's position in the many-body ground states and is directly related to the quantum metric of the occupied bands in band insulators. 
On the other hand, for correlated electron systems such as fractional Chern insulators, the ground state cannot even be approximated by any Slater determinant and the band picture does not apply. Nonetheless, the static structure factor and thus the quantum weight remain well-defined. 
Intuitively, the quantum weight represents quantum fluctuations in the center-of-mass position of electrons. A large quantum weight indicates that electrons in the system are delocalized even though the system is insulating. 

In the following sections, we relate the quantum weight to optical responses and many-body quantum geometry in general systems with short-range or Coulomb interactions and demonstrate that the quantum weight is a central quantity characterizing quantum many-body systems. 

\begin{widetext}
\section{Quantum weight, optical sum rule, and many-body quantum geometry} \label{sec:noint_sumrule}
\end{widetext}
In this section, we show how the quantum weight is related to the \red{density and optical} responses. We focus on systems with short-range interaction here and discuss Coulomb systems later. 

Since the structure factor and the quantum weight are defined through the density correlation, we can relate the quantum weight to the density response through the fluctuation-dissipation theorem.
Consider the density response of a solid to an external potential $V_{\rm ext}$ with wavevector $\vec{q}$ and frequency $\omega$. The induced density change is characterized by the density-density response function $\Pi(\vec{q},\omega)$: 
    $\rho(\vec{q},\omega) = \Pi(\vec{q},\omega) V_{\rm ext}(\vec{q},\omega)$. 
By the fluctuation-dissipation theorem~\cite{callen_irreversibility_1951}, $\Im\Pi(\vec{q},\omega)$ is directly related to the dynamical structure factor: 
   $ -\Im \Pi(\vec{q}, \omega) = e^2 S(\vec{q}, \omega)/(2\hbar)$, \label{eq:Pi-S}
where the dynamical structural factor is defined as 
  $  S(\vec{q},\omega) = (1/V)\int_{-\infty}^{\infty}\dd{t} e^{i\omega t} \expval{\hat{n}_{\vec{q}}(t)\hat{n}_{-\vec{q}}(0)}$. 
Integrating this equality over frequencies yields a sum rule that relates the dynamical density response to the static structure factor: 
\begin{eqnarray}
    -\int_0^{\infty}\dd\omega \Im \Pi(\vec{q}, \omega) = \frac{\pi e^2}{\hbar} S_{\vec{q}}, \label{eq:Pi-S_integrated_new}
\end{eqnarray} 
where we have used $S(\vec{q},\omega)=0$ for $\omega<0$ at zero temperature.

The density response $\Pi$ in the long wavelength limit is related to the optical conductivity $\sigma(\omega)$ at $q\rightarrow 0$. From the continuity equation $-i\omega\rho+i\vec{q}\vdot\vec{j}=0$, it follows that in the $\vec{q}\to 0$ limit,
\begin{align}
    \Pi(\vec{q}\to 0,\omega) &= -i\frac{q_{\alpha}q_{\beta}\sigma_{\alpha\beta}(\omega)}{\omega} \label{eq:Pi_sigma_noint}. 
\end{align}
From Eq.~\eqref{eq:Pi_sigma_noint} and Eq.~\eqref{eq:Pi-S_integrated_new},  we can directly relate the optical conductivity to the quantum weight as
\begin{align}
     \int_0^\infty \dd{\omega} \frac{\Re\sigma_{\alpha\alpha}(\omega)}{\omega} 
    & = \frac{e^ 2}{2\hbar} K_{\alpha\alpha}, 
    \label{eq:sigmaL_K2_new} 
\end{align}
\red{Here and hereafter, we assume that the system has principal axes so that $\Re\sigma_{\alpha\beta}\propto \delta_{\alpha\beta}$ and similar for other responses.}
This sum rule connects the negative first moment of optical conductivity to the ground-state static structure factor at small $\vec q$, and therefore can be regarded as a generalization of the $f$-sum rule, which relates the zeroth moment of optical conductivity (i.e., optical spectral weight) to the electron density. Eq.~\eqref{eq:sigmaL_K2_new} is finite when the DC longitudinal conductivity vanishes ($\sigma_{\alpha\alpha}(0)=0$), showing that the static structure factor in insulators at long wavelengths behaves as $q^2$, as mentioned earlier.

We now compare Eq.~\eqref{eq:sigmaL_K2_new} with the SWM sum rule for insulating states~\cite{souza_polarization_2000}, which relates the optical conductivity $\sigma(\omega)$ to the many-body quantum metric: 
\begin{align}
     \int_0^\infty \dd{\omega}\frac{\Re \sigma_{\alpha\alpha}(\omega)}{\omega} 
    & = \frac{\pi e^ 2}{\hbar} G_{\alpha\alpha}. \label{eq:SWM} 
\end{align}
where the many-body quantum metric $G$ is defined by the change of the many-body ground state under twisted boundary condition:  %
\begin{align}
    G_{\alpha\beta} &\equiv \frac{1}{V}\Re\mel{\partial_{\kappa_\alpha}\Psi_{\vec{\kappa}}}{(1-P_{\vec{\kappa}})}{\partial_{\kappa_\beta}\Psi_{\vec{\kappa}}}_{\vec{\kappa}=0}, \label{eq:mbmetric}
\end{align}
where $\ket{\Psi_{\kappa}}$ is the ground state satisfying the twisted boundary condition specified by $\vec{\kappa}$: $\Psi_{\vec{\kappa}}(\vec{r}_1, \dots, \vec{r}_n + \vec{L}_{\mu}, \dots, \vec{r}_N) =  e^{i\vec{\kappa}\vdot\vec{L}_{\mu}}\Psi_{\vec{\kappa}}(\vec{r}_1, \dots, \vec{r}_n, \dots, \vec{r}_N)$ with $\vec{L}_{\mu}$ the vector of system size in $\mu$-direction, and  $P_{\vec{\kappa}}=\op{\Psi_{\vec{\kappa}}}$. $\kappa=0$ corresponds to the periodic boundary condition. 

The SWM sum rule~\eqref{eq:SWM} can be derived from the standard Kubo formula for the optical conductivity~\cite{souza_polarization_2000}:
\begin{align}
    &\sigma_{\alpha\beta}(\omega) = \frac{e^2}{\hbar V} \sum_{n, m} \frac{-iE_{nm} A^{\alpha}_{nm}A^{\beta}_{mn} f_{nm}}{\hbar\omega + E_{nm} + i\delta}, \label{eq:interacting_sigma}
\end{align}
where $E_{nm}=E_n-E_m$ is the energy difference between $n$-th and $m$-th many body eigenstate, and $\partial_{\mu}$ is the derivative with respect to $\kappa_{\mu}$. $f_{nm}=f_n-f_m$ with the probability $f_n$ that $n$-th eigenstate is realized. At zero temperature, the canonical distribution gives $f_0 = 1$ where the state $n=0$ is the ground state and otherwise $f_n=0$.
$A^{\mu}_{nm}=\mel{n,\vec{\kappa}}{i\partial_{\mu}}{m,\vec{\kappa}}$ is the interband Berry connection for the $n$-th and $m$-th eigenstates the interacting system under the boundary condition $\vec{\kappa}$. The SWM sum rule~\eqref{eq:SWM} follows from Eq.~\eqref{eq:interacting_sigma} and a relation $G_{\alpha\beta} = (1/V)\sum_{n\neq 0}\Re[A^{\alpha}_{0n}A^{\beta}_{n0}]$.

It should be clear that the many-body quantum metric and the quantum weight are conceptually distinct quantities: the former describes the change of many-body ground states under twisted boundary conditions, while the latter describes the density correlation in the ground state.
Nonetheless, we conclude from Eqs.~\eqref{eq:sigmaL_K2_new} and ~\eqref{eq:SWM} that, provided that electron-electron interaction in the system is {\it not too long-ranged} (see section \ref{sec:wigner_solids} and \ref{sec:Quantum_weight_Coulomb}), the many-body quantum metric $G$ and quantum weight $K$ are equal in the thermodynamic limit:  
\begin{align}
    G_{\alpha\beta} = \frac{K_{\alpha\beta}}{2\pi} = \frac{1}{2}\eval{\pdv[2]{S_{\vec{q}}}{q_{\alpha}}{q_{\beta}}}_{\vec{q}\to 0}. %
    \label{eq:K=2piG}
\end{align}

One may verify the equality between $K$ and $G$ explicitly for noninteracting band insulators. 
In this case, the quantum weight is given by the integral of $g$ over the Brillouin zone as shown in Eq.~\eqref{K-BI}, which is indeed the many-body quantum metric of band insulators. This confirms the relation Eq.~\eqref{eq:K=2piG}. %

It should be noted that while the quantum weight $K$ is defined solely with a single ground state, the quantum metric $G$ is not; rather it is defined with the derivative of $\ket{\Psi_{\vec{\kappa}}}$ with respect to $\vec{\kappa}$. Indeed, extracting many-body quantum geometry from a {\it single} ground state is a nontrivial task. In previous works, the many-body quantum metric was discussed in relation to the polarization fluctuation in the ground state~\cite{souza_polarization_2000, resta_polarization_2006}. However, the definition of polarization operator in terms of electron position applies only to systems with open boundary conditions, and it is unclear whether the polarization fluctuation thus defined is a bulk quantity independent of microscopic details at the boundary.   %
In contrast, our result Eq.~\eqref{eq:K=2piG} provides a direct method for extracting the many-body quantum metric from the static structure factor---an equal-time correlation function that can be both numerically calculated from a single ground state wavefunction and experimentally measured. \red{Our approach offers a practical shortcut for computing the many-body quantum metric in correlated systems, as demonstrated in Ref.~\cite{zaklama_structure_2024}.} (We note that a general formula for many-body Chern number in terms of equal-time correlation function is still lacking.)

\begin{widetext}
\section{Topological bound on quantum weight} \label{sec:topological_bound}
\end{widetext}
Through the relation with many-body quantum geometry, we now show that the quantum weight of two-dimensional electron systems is generally related to their ground-state topology, in the form of an inequality:
\begin{align}
    K \ge \abs{C}. \label{eq:topo_bound}
\end{align}
This inequality for many-body quantum geometry was derived in our recent work~\cite{onishi_topological_2024} based on the positivity of optical absorption of circularly polarized light. Here we provide an alternative derivation in terms of many-body quantum geometry.

Many-body quantum geometry is characterized by the many-body quantum geometric tensor:
\begin{align}
    Q_{\alpha\beta} \equiv \frac{1}{V}\mel{\partial_{\kappa_\alpha}\Psi_{\vec{\kappa}}}{(1-P_{\vec{\kappa}})}{\partial_{\kappa_\beta}\Psi_{\vec{\kappa}}}_{\vec{\kappa}=0}.
\end{align}
The real part of $Q_{\alpha\beta}$ gives the many-body quantum metric $G_{\alpha\beta}$ (Eq.~\eqref{eq:mbmetric}), while its imaginary part gives the many-body Berry curvature $\Omega=-2\Im Q$. 

\red{From the properties of many-body quantum geometric tensor, we can derive an inequality between the quantum metric $G_{\alpha\beta}$ and the Berry curvature $\Omega_{xy}$:}  
\begin{align}
    G_{xx} + G_{yy} \ge \abs{\Omega_{xy}}. \label{eq:geometric_ineq}
\end{align}
This inequality follows from the fact that the many-body quantum geometric tensor $Q$ is always semi-positive: for any complex vector $\vec{v}=(v_x, v_y)$, 
\begin{align}
    \vec{v}^{\dagger} Q\vec{v} \ge 0. \label{eq:Q_positive}
\end{align} 
This readily follows by rewriting the left-hand side as $\vec{v}^{\dagger} Q\vec{v} = \norm{(1-P)(\vec{v}\vdot\nabla_{\vec{\kappa}})\ket{\Psi}}^2$. In particular, choosing $\vec{v}=(1, \pm i)$ reduces Eq.~\eqref{eq:Q_positive} to $G_{xx}+G_{yy} \pm \Omega_{xy} \ge 0$, which is equivalent to Eq.~\eqref{eq:geometric_ineq}. This mathematical inequality for many-body quantum geometry was derived in Ref.~\cite{ozawa_relations_2021}. A similar inequality for the quantum geometric tensor of band insulators was originally derived by Roy~\cite{roy_band_2014}, see also~\cite{peotta_superfluidity_2015}. 

Physically, we can understand the geometric inequality~\eqref{eq:geometric_ineq} as a consequence of the positivity of the optical absorption. In fact, the quantum geometric tensor is related to the optical conductivity through the following sum rule~\cite{onishi_fundamental_2024}:
\begin{align}
    \int_0^\infty \dd{\omega}\frac{\sigma^{\rm abs}_{\alpha\beta}(\omega)}{\omega} 
    & = \frac{\pi e^ 2}{\hbar} Q_{\alpha\beta}. \label{eq:sum_rule_Q} 
\end{align}
where $\sigma^{\rm abs}(\omega)=(\sigma(\omega)+\sigma^{\dagger}(\omega))/2$ is the hermitian part of the conductivity and represents the optical absorption as discussed in Ref.~\cite{onishi_fundamental_2024}. This sum rule follows from Eq.~\eqref{eq:interacting_sigma} and a relation $Q_{\alpha\beta} = \sum_{n\neq 0}A^{\alpha}_{0n}A^{\beta}_{n0}$. The real part of Eq.~\eqref{eq:sum_rule_Q} reduces to the SWM sum rule~\eqref{eq:SWM}, while its imaginary part reduces to the Kramers-Kronig relation of the Hall conductivity. 
Since the optical absorption is semi-positive, it follows $\Re\sigma_{xx}+\Re\sigma_{yy}\ge \abs{\Im \sigma_{xy}}$. Combining this inequality with the sum rule~\eqref{eq:sum_rule_Q}, we obtain Eq.~\eqref{eq:geometric_ineq}, linking the geometric inequality~\eqref{eq:geometric_ineq} to the positivity of the optical absorption.

In the thermodynamic limit, the many-body Berry curvature becomes independent of the choice of twisted boundary condition $\kappa$ and reduces to the many-body Chern number as $2\pi\Omega_{xy} =C$, as shown by Niu, Thouless and Wu~\cite{niu_quantized_1985} and further elaborated in Ref.~\cite{ceresoli_orbital_2007} for noninteracting cases and then later in Ref.~\cite{kudo_many-body_2019} for interacting cases. Therefore, combining Eq.~\eqref{eq:geometric_ineq} and Eq.~\eqref{eq:K=2piG} yield the topological bound on the quantum weight~\eqref{eq:topo_bound}. Our bound also applies to fractional quantum Hall states and fractional Chern insulators~\cite{onishi_topological_2024}. This extension is achieved by generalizing the many-body quantum geometric tensor to the non-Abelian case to handle degenerate ground states~\cite{onishi_fundamental_2024}.   

\begin{widetext}
\section{Structure factor and optical response of a Coulomb solid} \label{sec:wigner_solids}
\end{widetext}

We now turn to electron systems with long-range Coulomb interaction. We shall show that (i) the structure factor of three-dimensional (3D) solids, i.e., real bulk materials, is significantly affected by the long-range nature of the Coulomb interaction, and  (ii) the optical sum rule (Eq.~\eqref{eq:sigmaL_K2_new}) does {\it not} hold.   
Before presenting a general theory, \red{we first consider an illustrative example of 3D Coulomb solids in this section.}

Consider a 3D electron system where Coulomb interaction and periodic potential dominate over the kinetic energy, leading to a periodic lattice of strongly localized electrons~\cite{onishi_universal_2024, bagchi_dielectric_1969, dolgov_admissible_1981, iwamoto_sum_1984}. By treating the kinetic energy perturbatively around the classical limit, the long-wavelength fluctuations can be described in terms of the displacement from the mean position of electrons. For the simplest case of one electron per unit cell (a Wigner solid), the effective Lagrangian takes the following form:
\begin{align}
    {\cal L} &= \sum_{\vec q} \frac{m}{2}\qty(\abs{\dot{\vec{u}}_{\vec{q}}}^2 - (\omega_0^2 + \omega_p^2 \frac{q_\alpha q_\beta}{q^2}) u_{\vec{q}\alpha}u_{-\vec{q}\beta}), \label{eq:Lag}
\end{align}
where $u_{\vec{q}}=(1/\sqrt{N})\sum_i e^{-i\vec{q}\vdot\vec{R}_i}\vec{u}_i$ is the Fourier transform of the displacement $\vec{u}_i$ of the $i$-th electron from its mean position $\vec{R}_i$, $\omega_0$ represents the pinning of electron lattice by the potential, and the last term with the bare plasma frequency $\omega_p \equiv (ne^2/(m\epsilon_0))^{1/2}$ describes the increase in Coulomb interaction energy due to the change in charge density caused by electron displacements: $\rho = - e {\bf \nabla} \cdot {\vec u}$. Here, $n, m$ are the electron density and the free electron mass, respectively.

The low-energy spectrum of this system consists of a longitudinal mode 
($\vec{u}_{\vec{q}}^{L}\parallel\vec{q}$) and two transverse modes ($\vec{u}_{\vec{q}}^{T}\perp \vec{q}$). Importantly, the dispersion relations of longitudinal and transverse modes differ even at $\vec{q}\to 0$. This is because the longitudinal mode, which represents the plasma oscillation, involves electron density fluctuation and therefore costs finite Coulomb energy even at $\vec{q}\to0$ due to the long-range nature of Coulomb force. Indeed, it is clear from Eq.~\eqref{eq:Lag} that the frequency of the transverse mode is $\omega_T=\omega_0$, while that of the longitudinal mode is $\omega_L=\sqrt{\omega_0^2 + \omega_p^2}>\omega_T$. Here, $\omega_L$ is the real plasma frequency of the system, which is larger than $\omega_p$ due to the presence of periodic potential. 

We now calculate the quantum weight for this system by quantizing the  Lagrangian \eqref{eq:Lag} (also see Supplemental Materials for details). 
The number density operator $n(\vec{r})$ is given by 
$n(\vec{r}) = \sum_{i} \delta(\vec{r}-(\vec{R}_i + \vec{u}_i))$
and thus its Fourier transform $n_{\vec{q}}$ is given by 
\begin{align}
    n_{\vec{q}} &= \int\dd{\vec{r}} e^{-i\vec{q}\vdot\vec{r}} n(\vec{r}) = \sum_{i} e^{-i\vec{q}\vdot(\vec{R}_i+\vec{u}_i)}
\end{align}
Therefore, $S_{\vec{q}}$ at small $q$ is given by
\begin{align}
    S_{\vec{q}} &= n\expval{\qty(\vec{q}\vdot\vec{u}_{\vec{q}})\qty(\vec{q}\vdot\vec{u}_{-\vec{q}})}, 
    \; {\rm for} \; {\vec q} \rightarrow 0. \label{eq:Sq_wigner}
\end{align}
Importantly, only the longitudinal mode ($\vec{u}_{\vec{q}}^{L}\parallel \vec{q}$) contributes to $S_{\vec{q}}$. 
From the amplitude fluctuation of a harmonic oscillator $\langle|\vec{u}_{\vec{q}}^{L}|^2\rangle = \hbar/(2m\omega_L)$, we find the quantum weight as
\begin{align}
    K & = \frac{\pi n \hbar}{m\omega_L} = \frac{\pi\hbar\epsilon_0}{e^2}\frac{\omega_p^2}{\sqrt{\omega_p^2 + \omega_0^2}}. \label{eq:K_wigner}
\end{align}
For a given electron density, as the potential depth decreases, $\omega_0$ and therefore $\omega_L$ decrease, leading to an increase of the quantum weight, consistent with electrons becoming less localized and, therefore, more quantum. 

To see how the quantum weight is related to optical responses, we also calculate the optical conductivity.
The electric current at finite $\vec{q}$ is $\vec{j}_{\vec{q}}=e\sqrt{N}\sum_i\dot{\vec{u}}_{\vec{q}}$ up to leading order in $\vec{q}$. Noting that the optical field is always transverse ($\vec{E} \perp \vec{q}$) and thus couples only to the transverse mode, we find the optical conductivity as (with $\delta \to +0)$
\begin{align}
    \sigma(\omega) = -i\omega \epsilon_0\frac{\omega_p^2}{\omega_T^2-(\omega+i\delta)^2}, \label{eq:sigma_wigner}
\end{align}
which has a pole at $\omega_T$. 

We note that the same result for $\sigma(\omega)$ is obtained by considering the current in response to a longitudinal electric field $\vec{E} \parallel \vec{q}$.  However, it is crucial to recognize that, in this case, a longitudinal external field induces a density modulation $\rho$, leading to dielectric screening due to the long-range Coulomb interaction. As a result, in 3D Coulomb systems, the {\it total} electric field $\vec{E}$ differs from the external field $\vec{E}_{\rm ext} \parallel \vec{q}$ even at $q\rightarrow 0$, and their ratio defines the dielectric constant: $\vec{E}_{\rm ext} = \epsilon \vec{E}$.   
The current response to the total field $\vec{E} \parallel \vec{q}$ in the long wavelength limit, which defines the conductivity as physically measured, gives the same expression Eq.~\eqref{eq:sigma_wigner} as calculated from the transverse field $\vec{E} \perp \vec{q}$. (See Supplemental Materials and Ref.~\cite{onishi_universal_2024} for more details.)

From the optical conductivity, we obtain its negative-first moment:
\begin{align}
    & \int_0^\infty \dd{\omega} \frac{\Re\sigma_{\alpha\alpha}(\omega)}{\omega} = \frac{\hbar\epsilon_0}{2e^2}\frac{\omega_p^2}{\omega_T}. \label{eq:G_wigner}
\end{align}
Importantly, comparison with the expression for the quantum weight~\eqref{eq:K_wigner} shows that Eq.~\eqref{eq:sigmaL_K2_new} no longer holds since longitudinal and transverse modes have different frequencies at ${\vec q} \rightarrow 0$: $\omega^L \neq \omega^T$. In general, the sum rule~\eqref{eq:sigmaL_K2_new} relating optical conductivity to structure factor (and therefore quantum weight) does \textit{not} hold in 3D Coulomb systems due to the long-range nature of the Coulomb interaction. \red{The long-range Coulomb interaction alters the behavior of the longitudinal mode, making it distinct from the transverse mode even at $\vec{q}\to 0$. Consequently, the structure factor, which is associated with the longitudinal modes, is no longer related to the optical conductivity through Eq.~\eqref{eq:sigmaL_K2_new}, as the latter is associated with the transverse modes. } 

\red{Note that the fluctuation of electron displacement $\langle|\vec{u}_{\vec{q}}^{L}|^2\rangle$, and thus the quantum weight $K$, can also be regarded as polarization fluctuation~\cite{resta_electron_1999, souza_polarization_2000}. Then our analysis of  Coulomb solid shows explicitly that for real bulk solids, the polarization fluctuation is {\it different} from the many-body quantum metric because of dielectric screening in three dimensions; the crucial distinction between the two was missed in~\cite{souza_polarization_2000}.} In the next section, we will discuss the general relation between the quantum weight and the optical responses in Coulomb systems.

\begin{widetext}
\section{Static and Optical Quantum weights in general Coulomb systems} \label{sec:Quantum_weight_Coulomb}
\end{widetext}

As shown explicitly in the previous section, the negative-first moment of optical conductivity and quantum weight are \textit{not} equal in 3D Coulomb systems. 
To distinguish them, we shall refer to the quantum weight defined through the static structure factor as \textit{static} quantum weight, and the negative-first moment of optical conductivity as \textit{optical} quantum weight---in analogy with the zero-th moment of optical conductivity being called optical spectral weight. 
While the optical sum rule~\eqref{eq:sigmaL_K2_new} shows the static and optical quantum weight are equal for systems with short-range interaction, it is not the case in 3D Coulomb systems. Instead of Eq.~\eqref{eq:sigmaL_K2_new}, another sum rule relates the energy loss function to the static quantum weight, as we describe below.

To begin with, let us reexamine the relation~\eqref{eq:Pi_sigma_noint} between the density response $\Pi$ and the optical response $\sigma$ in Coulomb systems. 
It is important to note that density fluctuation is accompanied by longitudinal current only, i.e., the current parallel to the wavevector ${\vec j}^L \parallel \vec{q}$. Thus, the conductivity for the longitudinal current $\sigma^L$ is generally related to the density response function $\Pi$ as 
\begin{align}
    \Pi(\vec{q},\omega) &= -i\frac{q_{\alpha}q_{\beta}\sigma^L_{\alpha\beta}(\vec{q},\omega)}{\omega\epsilon(\vec{q},\omega)} \label{eq:Pi_sigma_new}. 
\end{align}
Compared to the previous relation~\eqref{eq:Pi_sigma_noint}, a key addition is $\epsilon$ appearing in the denominator in Eq.~\eqref{eq:Pi_sigma_new}, which accounts for electric field screening due to density fluctuation. The reason is as follows.   The conductivity (as physically measured) 
$\sigma^L(\vec{q}, \omega)$ is defined as the longitudinal current $\vec{j}^L$ in response to the {\it total} electric field ${\vec E}$ parallel to the wavevector $\vec{q}$: $\vec{j}^L_{\vec{q}}(\omega) = \sigma^L(\vec{q},\omega) \vec{E}(\vec{q},\omega)$. On the other hand, $\Pi$ describes the density response to the \textit{external} potential.  
The external field ${\vec E}_{\rm ext} = -\grad V_{\rm ext}$ differs from the total electric field by the dielectric function: $\vec{E}=\vec{E}_{\rm ext}/\epsilon$, which explains the more general relation ~\eqref{eq:Pi_sigma_noint}.

In systems with short-range interactions or in 1D or 2D Coulomb systems, electric field screening does not occur at $\vec{q}\to 0$, and thus $\epsilon(\omega)=1$, reducing Eq.~\eqref{eq:Pi_sigma_new} to Eq.~\eqref{eq:Pi_sigma_noint} at the leading order in $\vec{q}$.  
However, for 3D systems with long-range Coulomb interaction $U(\vec{q})=1/(\epsilon_0 q^2)$, due to dielectric screening we expect $\epsilon(\omega)\neq 1$ even at $\vec{q}\to 0$.  
Indeed, in the most general form, the density response function $\Pi$ is related to the dielectric function $\epsilon(\omega)$ as (see Ref.~\cite{pines2018theory} and Supplemental Materials)
\begin{align}
    \epsilon(\vec{q},\omega) &= (1+U(\vec{q})\Pi(\vec{q},\omega))^{-1} \label{eq:eps-Pi_new}
\end{align}
Typically, $\Pi(\vec{q},\omega)\propto q^2$ at $\vec{q}\to 0$ in the gapped systems. Thus, provided that $U(q)$ is not too singular, we have $U(\vec{q})q^2 \to 0$ at $\vec{q}\to 0$, resulting in $\epsilon(\omega)=1$. However, this is not the case for 3D Coulomb interaction where $U(\vec{q})=1/(\epsilon_0 q^2)$, leading to $\epsilon(\omega)\neq 1$ even at $\vec{q}\to 0$. This reflects the electric field screening effect due to the long-range nature of Coulomb interaction.

By plugging a general relation~\eqref{eq:Pi_sigma_new} into Eq.~\eqref{eq:Pi-S_integrated_new}, we find a general sum rule for the static quantum weight $K$:
\begin{align}
     \int_0^\infty \dd{\omega} \frac{1}{\omega} \Re\qty[\frac{\sigma_{\alpha\alpha}(\omega)}{\epsilon_{\alpha\alpha}(\omega)}] 
    & = \frac{e^ 2}{2\hbar} K_{\alpha\alpha}. \label{eq:sigmaL_K_new} 
\end{align}
Here we removed the superscript $L$, because the physical conductivity $\sigma^L$ at $\vec{q}\to0$ does not depend on whether the current is longitudinal or not. 

Our sum rule Eq.~\eqref{eq:sigmaL_K_new} relates the static quantum weight to the conductivity and dielectric function, %
which applies to general many-body systems and includes the optical sum rule ~\eqref{eq:sigmaL_K2_new} as a special case with $\epsilon=1$. 

Specifically for 3D Coulomb systems, 
we can further simplify Eq.~\eqref{eq:sigmaL_K_new} by relating the optical conductivity to the dielectric function. Note that the total electric field $\vec{E}=\vec{E}_{\rm ext}/\epsilon$ is related to the polarization $\vec{P}$ and the external field $\vec{E}_{\rm ext}$ as $E_{\rm ext}=\vec{E}+\vec{P}/\epsilon_0$. Then, since $\dot{\vec{P}}=\vec{j}=\sigma \vec{E}_{\rm tot}$, we have $\sigma(\omega)=-i\omega\epsilon_0(\epsilon(\omega)-1)$. 
Plugging this relation into Eq.~\eqref{eq:sigmaL_K_new} yields a sum rule that relates the static quantum weight of 3D Coulomb systems to the {\it inverse} dielectric function (see, for example, Ref.~\cite{pines2018theory}): 
\begin{align}
     \int_0^\infty \dd{\omega}\Im\qty[-\frac{1}{\epsilon_{\alpha\alpha}(\omega)}] 
     &= \frac{1}{2\hbar} \frac{e^ 2  K_{\alpha\alpha}}{ \epsilon_0}, & \nonumber \\
	&\text{when $U(\vec{q})=1/(\epsilon_0 q^2)$.}& \label{eq:loss-func_K1_new} 
\end{align}
The integrand of the left-hand side of Eq.~\eqref{eq:loss-func_K1_new} is called energy loss function and represents density excitations such as plasmon. It can be directly measured by the inelastic X-ray scattering or the electron energy loss spectroscopy~\cite{schulke_dynamic_1995, caliebe_dynamic_2000, abbamonte_dynamical_2008}. Therefore, the sum rule~\eqref{eq:loss-func_K1_new} provides a direct method of determining the quantum weight of real materials.

We can verify Eq.~\eqref{eq:loss-func_K1_new} in the Wigner solid example introduced in the previous section. Since the charge density at small $\vec{q}$ is given by $\rho_{\vec{q}}=e\sqrt{N}(-i\vec{q}\vdot\vec{u}_{\vec{q}})$, the density response function $\Pi(\vec{q}, \omega)$ is given by the zero-point fluctuation in $\vec{u}_{\vec{q}}$:  
\begin{align}
    \Pi(\vec{q}\to 0,\omega) &= - \epsilon_0 q^2 \frac{\omega_p^2}{\omega_L^2-(\omega+i\delta)^2}
\end{align}
with $\omega_p^2=ne^2/(m\epsilon_0)$ the plasma frequency, $n$ the electron density, and $\delta\to +0$ is assumed. 
Note that since $\Pi$ probes the longitudinal modes $u^L$, $\Pi$ has a pole at $\omega_L$. 
From $\Pi$, we can also obtain the dielectric function (see Supplemental Materials) as~\cite{bagchi_dielectric_1969, dolgov_admissible_1981}
\begin{align}
    \epsilon(\omega) = \frac{\omega_L^2 - (\omega+i\delta)^2}{\omega_T^2 - (\omega+i\delta)^2}. \label{eq:eps}
\end{align}
Note that the pole of $1/\epsilon$ corresponds to the longitudinal mode, while the pole of $\epsilon$ corresponds to the transverse mode. Plugging this into the sum rule~\eqref{eq:loss-func_K1_new}, we find $K=\pi\hbar\epsilon_0 \omega_p^2/(e^2\omega_L)$, which coincides with Eq.~\eqref{eq:K_wigner} obtained directly from the static structure factor and thus verifies the sum rule~\eqref{eq:loss-func_K1_new}.

The concept of quantum weight we defined in terms of static structure factor also provides a useful perspective on theory of electron localization in insulating systems, initiated by Kohn~\cite{kohn_theory_1964} 
and further developed over decades~\cite{resta_electron_1999, souza_polarization_2000}. \red{To characterize insulating states in general,  Ref.~\cite{resta_electron_1999} introduced a quantity called ``localization tensor'' defined as $N \log |\langle e^{i2\pi X/L} \rangle| $, and 
Ref.~\cite{souza_polarization_2000} introduced polarization fluctuation $(1/V^2)\expval{X_{\alpha}X_{\beta}}$,  where $\vec{X}\equiv\sum_i \vec{x}_i$ is the electrons' center of mass position operator, and $L$ and $V$ denote the linear size and the volume of the system respectively. Both quantities characterize quantum fluctuation in electron's center of mass. 
}

\red{However, the localization tensor involves multi-body correlations between all particles, as opposed to two-body correlation. Polarization fluctuation also has subtle issues due to the fact that the position operator is not well-defined for systems with periodic boundary conditions~\cite {resta_insulating_2011}.} This was pointed out by Resta in his insightful work~\cite{resta_polarization_2006}, where it was shown that polarization fluctuation depends on the boundary condition, and for 3D systems with long-range Coulomb interaction, the SWM relation holds only for the ``purely transverse'' boundary condition.

Instead of polarization fluctuation, we consider long-wavelength density fluctuation described by the static structure factor $S_{\vec q}$, which is a bulk property and a physical observable. 
Heuristically, the connection between the two approaches can be made if we identify the polarization $\vec{P}$ as the electrons' center of mass position $\vec{X}$ and use the relation between polarization and density $\nabla\vdot\vec{P}=-\rho$. %
\red{By using the structure factor instead of center-of-mass position, we define the quantum weight that is independent of the boundary condition, and directly relate it to density response via the dielectric sum rule for general systems (see Eqs.~\eqref{eq:loss-func_K1_new} and \eqref{eq:sigmaL_K_new}). This approach provides a natural way to measure them experimentally, as well as to compute numerically without boundary condition subtleties.
Moreover, our approach clarifies the difference between dielectric and optical responses and therefore the difference between static and optical quantum weights in 3D Coulomb systems. We also identify the condition under which the SWM relation between polarization fluctuation and many-body quantum metric holds}.

We note that the dielectric function we discussed is a quantity that macroscopically relates the total field to the external field on the length scale much larger than the lattice constant. Throughout this paper, we implicitly assume the wavevector of the external perturbation $\vec{q}$ is sufficiently small compared to the reciprocal lattice vectors $\vec{G}$ so that physical responses at wavevector $\vec{q}+\vec{G}$ are negligibly small~\cite{dolgov_admissible_1981}.

\begin{widetext}
\section{Bounds on static and optical quantum weights} \label{sec:bound}
\end{widetext}
Real materials are obviously more complicated than Wigner solid. %
In order to estimate the static quantum weight of real materials, we now derive rigorous upper and lower bounds on this quantity based on its relation to the inverse dielectric function, Eq.~\eqref{eq:loss-func_K1_new}.
To this end, it is convenient to define the $i$-th moment of the energy loss function:
\begin{align}
    K_i \equiv \int_0^{\infty}\dd{\omega}\omega^i\Im\qty[-\frac{1}{\epsilon_{\alpha\alpha}(\omega)}]
\end{align}
In particular, we focus on $K_0, K_{-1}$, and $K_{1}$. $K_0 = e^2 K/(2\hbar \epsilon_0)$ is simply the quantum weight up to a factor as seen from Eq.~\eqref{eq:loss-func_K1_new}, while $K_1$ and $K_{-1}$ are related to the bare plasma frequency and static dielectric constant respectively:
\begin{align}
    &K_1 = \frac{\pi \omega_p^2}{2}, \label{eq:f-sum_loss_func} \\
    &K_{-1} = \frac{\pi}{2} (1-\epsilon_{\alpha\alpha}^{-1}(0)), \label{eq:KK_loss_func}
\end{align}
Eq.~\eqref{eq:f-sum_loss_func} is the $f$-sum rule~\cite{pines2018theory, Mahan2000}, while Eq.~\eqref{eq:KK_loss_func} is a consequence of the Kramers-Kronig relation for $1/\epsilon$ which describes the density response~\cite{pines2018theory, landau2013electrodynamics}. 

Bounds on the quantum weight can be obtained by exploring the inequality relations between $K_{-1}$, $K_0$ and $K_1$.  
First,  the Cauchy-Schwartz inequality, $\int_0^{\infty}\abs{f(x)}^2\dd{x}\int_0^{\infty}\abs{g(x)}^2\dd{x} \ge \abs{\int_0^{\infty} f(x)g(x)}^2$, with $f(x) = \sqrt{\omega\Im[-\epsilon^{-1}]}$, $g(x)=\sqrt{\omega^{-1}\Im[-\epsilon^{-1}]}$, yields an upper bound on $K_0$:  
\begin{align}
    K_0 \le \sqrt{K_{1} K_{-1}}. \label{eq:K0_upperbound}
\end{align}

To find the lower bound on $K_0$, note that the energy loss function $\Im[-1/\epsilon(\omega)]$ describes the energy absorption by density excitations and is semi-positive. For gapped systems, $\Im[-1/\epsilon(\omega)]$ can be finite only when $\hbar\omega\ge E_g$, with $E_g$ the energy of the lowest density excitation which couples to a long-wavelength potential, i.e., the lowest plasmon energy. Therefore, a lower bound on $K_0$ is obtained by replacing $\omega$ in $K_{-1}$ with $E_g/\hbar$:%
\begin{align}
     K_0 \ge \frac{E_g}{\hbar}K_{-1}.  \label{eq:K0_lowerbound}
\end{align}
Combining Eqs.~\eqref{eq:K0_upperbound} and \eqref{eq:K0_lowerbound} with the sum rules~\eqref{eq:f-sum_loss_func} and \eqref{eq:KK_loss_func}, we obtain an upper and a lower bound on the quantum weight in terms of electron density, static dielectric constant and the plasmon energy,  
\begin{align}
    \pi(1-\epsilon^{-1}) E_g \le e^2 K /\epsilon_0 \le \pi\sqrt{1-\epsilon^{-1}} \hbar\omega_p. \label{eq:bound_K}
\end{align}

It is important to note that the upper bound on $K_0$ given here is necessarily finite, because the static dielectric constant $\epsilon$ (in the limit $q\rightarrow 0$) is always greater than 1. Returning to the discussion after Eq.~\eqref{eq:sigmaL_K_new}, the boundedness of $K_0$ guarantees that the static structure factor of 3D Coulomb systems is indeed proportional to $q^2$. Also note that these bounds are all saturated when \red{the dielectric responses are exhausted by a single mode, i.e., when the single-mode approximation is exact}. Our bounds in the form of exact inequalities provide the rigorous ground for this approximation.

\begin{figure}[t]
    \centering
    \includegraphics[width=\columnwidth]{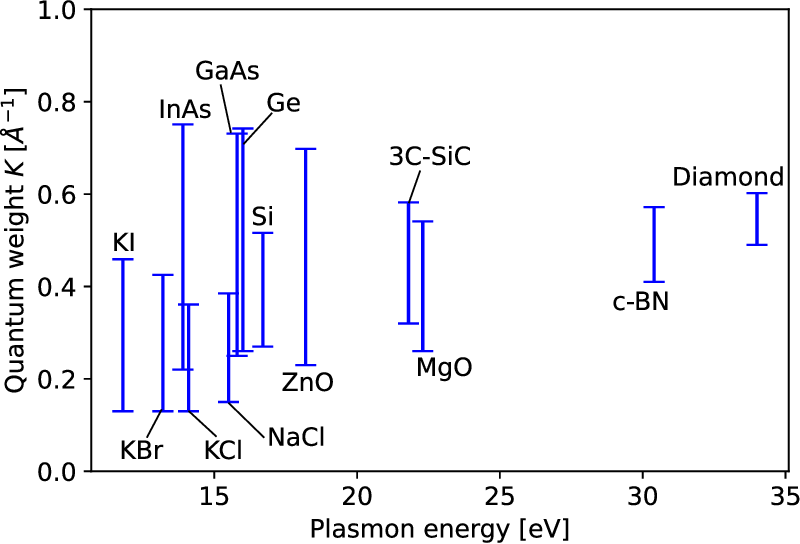}
    \caption{The bounds on the quantum weight $K$ for real materials. As the energy gap, we used the plasmon energy obtained from electron energy loss spectroscopy. The details of the used parameters are given in supplemental materials (also see Ref.~\cite{onishi_universal_2024}).}
    \label{fig:Kbound_CS}
\end{figure}

Remarkably, we find the bounds on the quantum weight~\eqref{eq:bound_K} work very well for real materials. \red{Assuming the electron-phonon coupling is negligible and} using the electron density and the measured values of static electronic dielectric constant and plasmon energy, we calculated the upper and the lower bound on the quantum weight. The results are shown in Fig.~\ref{fig:Kbound_CS}. The most remarkable case is diamond, where the quantum weight is bounded as $\SI{0.49}{\per\angstrom}\le K\le\SI{0.60}{\per\angstrom}$. The bounds also work well for cubic boron nitride (c-BN):  $\SI{0.41}{\per\angstrom}\le K\le \SI{0.57}{\per\angstrom}$. For all the materials we calculated, the upper bound is less than \SI{1}{\per\angstrom}. It is remarkable that the quantum weight of real materials lies within such a narrow range. %

Upper and lower bounds on the {\it optical} quantum weight can be obtained from optical sum rules similarly. We define the $i$-th moment of optical conductivity $\Re\sigma(\omega)=-\omega\Im\epsilon(\omega)$~\cite{onishi_fundamental_2024, onishi_universal_2024}:
\begin{align}
    W_i \equiv \int_0^{\infty}\dd{\omega}\omega^i \Re\sigma_{\alpha\alpha}(\omega).
\end{align}
$(2\hbar/e^2)W_{-1}$ is the optical quantum weight, while $W_0$ and $W_{-2}$ are given by
\begin{align}
    W_0 &= \frac{\pi}{2} \epsilon_0 \omega_p^2, \label{eq:W0} \\
    W_{-2} &= \frac{\pi}{2}\epsilon_0(\epsilon_{\alpha\alpha}(0)-1). \label{eq:W-2}
\end{align}
Eq.~\eqref{eq:W0} follows from the standard $f$-sum rule and Eq.~\eqref{eq:W-2} follows from Kramers-Kronig relation combined with a relation $\sigma(\omega)=-i\omega\epsilon_0(\epsilon(\omega)-1)$.
With these, we can derive a series of inequalities in the same way as we derive Eq.~\eqref{eq:K0_upperbound}, \eqref{eq:K0_lowerbound}. In particular, we can show the following upper and lower bounds on $W_{-1}$:
\begin{align}
    W_{-1} &\le \sqrt{W_0 W_{-2}} \le \frac{\hbar}{E_o}W_0, \\
    W_{-1} &\ge \frac{E_o}{\hbar} W_{-2}.
\end{align}
where $E_o$ is the optical gap, i.e., the energy of the lowest optical excitation (rather than the plasmon energy). 
Further using the standard sum rule $W_0=\pi \epsilon_0 \omega_p^2/2$ and Kramers-Kronig relation, it can be shown that 
\begin{align}
    \pi(\epsilon_{xx}-1)E_{o} \le \frac{2\hbar}{\epsilon_0}\int_0^{\infty}\dd{\omega}\frac{\Re \sigma_{xx}}{\omega} 
    \nonumber \\
    \le \pi\sqrt{\epsilon_{xx}-1}\hbar\omega_p \le \frac{\pi (\hbar\omega_p)^2}{E_o}.
\end{align}
where $E_o$ is the optical gap, i.e., the energy of the lowest optical excitation (rather than the plasmon energy). 

We note that the upper bound in terms of the spectral weight and optical gap was derived in Ref.~\cite{souza_polarization_2000} and compared to the calculated quantum metric for real materials in Ref.~\cite{sgiarovello_electron_2001}. The lower bound in terms of the optical gap and the static dielectric constant was derived in an earlier version of this work~\cite{onishi_quantum_2024}; see also related discussions in Refs.~\cite{aebischer_dielectric_2001, komissarov_quantum_2024}.
Recently, the tighter upper bound in terms of the spectral weight and static dielectric constant was independently found in Refs.~\cite{verma_instantaneous_2024, souza_optical_2024} using the Cauchy-Schwartz inequality.  
\begin{figure}[t]
    \centering
    \includegraphics[width=\columnwidth]{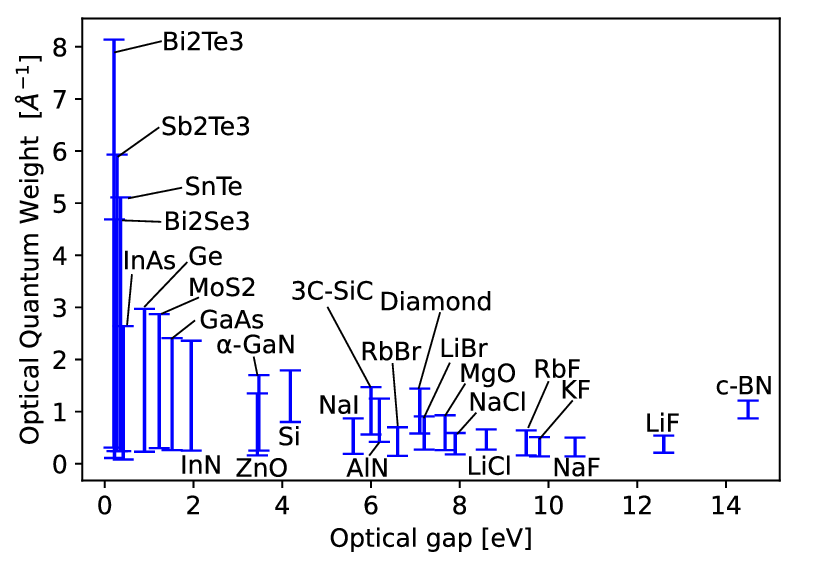}
    \caption{The bound on the optical quantum weight $(2\hbar/e^2)W_{-1}$ for real materials. As the energy gap, we used the optical gap obtained from optical measurement. The details of the used parameters are given in supplemental materials (also see Ref.~\cite{onishi_universal_2024}).}
    \label{fig:Gbound_CS}
\end{figure}

In general, the expression for quantum weight in terms of electron density and a single plasmon frequency $\omega_L$,  Eq.~\eqref{eq:K_wigner}, holds (approximately) in systems where the density response at long wavelength is (nearly) saturated by this single plasmon mode. Such single mode approximation was widely used in the study of liquid Helium~\cite{feynman_atomic_1954} and fractional quantum Hall systems~\cite{girvin_magneto-roton_1986}. As our analysis has shown, this approximation is exact for Wigner solid in the semiclassical limit ($\hbar \rightarrow 0$). It is also exact in a Coulomb gas without external potential, which corresponds to the limit $\omega_0\to 0$ in our discussion here. The quantum weight of 3D Coulomb gas, which we denote as $K_p$, is simply determined by the bare plasma frequency:  $e^2 K_p/\epsilon_0 = \pi \hbar \omega_p$. More generally, within the single mode approximation, the quantum weight in dimensionless units can be written compactly as: 
$K/ K_p  \approx \omega_p/\omega_L$.

We calculate the bounds on the optical quantum weight for real materials and the results are given in Fig.~\ref{fig:Gbound_CS}. For the materials with a large optical gap, the bounds constrain the value of the optical quantum weight to a quite narrow range. For example, in cubic boron nitride (c-BN), the optical quantum weight is bounded as $\SI{0.87}{\per\angstrom}\le 2\hbar/e^2)W_{-1} \le \SI{1.21}{\per\angstrom}$. The bounds also work well for some ionic crystals: for example, the bound for NaCl is $\SI{0.18}{\per\angstrom}\le (2\hbar/e^2)W_{-1} \le \SI{0.59}{\per\angstrom}$.

\red{Lastly, we comment on the effects of phonons. Throughout this paper, we have assumed that phonon contributions to dielectric and optical responses are negligible. Thus, the energy gap we refer to corresponds to electronic excitations rather than the many-body gap of the entire system, which remains gapless due to phonons. Investigating how phonons influence the quantum weight and the bounds derived in this work is beyond the scope of the present study and is left for future study.}

\section{Outlook} \label{sec:entanglement}
To summarize, by considering the static structure factor at long wavelengths, we introduced the general concept of quantum weight for many-body systems. Interestingly, besides describing long-wavelength charge fluctuations in the ground state, the quantum weight is intimately related by sum rules to dynamical responses including dielectric function $\epsilon(\omega)$ and optical conductivity $\sigma(\omega)$. We further relate the quantum weight to the ground-state topology and geometry for quantum many-body systems in general (provided that the interaction is not too long-ranged). The quantum weight brings together various aspects of quantum materials, highlighting the need for further theoretical and experimental studies.

\red{Very recently, an analysis of inelastic X-ray scattering data has been performed to determine the quantum weight of the ionic insulator LiF and compare its value with our theoretical bounds ~\cite{balut_quantum_2024}. Their work demonstrated that our theoretical proposal to measure quantum weight is indeed feasible and offers a powerful tool to study real quantum materials. Extending quantum weight measurements to a broader range of systems remains an important goal. 

As shown in Fig.~\ref{fig:Kbound_CS}, the quantum weights of the bulk materials we studied are all below \SI{1.0}{\per\angstrom}. Correspondingly, the dimensionless parameter $Ka$ with $a$ the lattice constant, remains of order 1. For example, the upper bound on the quantum weight of InAs is $Ka=4.5$. 
In contrast, large quantum weight can be found in narrow-gap semiconductors in low dimensions, since the quantum weight diverges at the gap-closing transition~\cite{onishi_fundamental_2024}. For example, recent first-principles calculations show that the antiferromagnetic topological insulator MnBi$_2$Te$_4$ multilayers has a quantum weight of $K>70$~\cite{ghosh_probing_2024}. 
Optical measurements of this material have also been reported~\cite{bac_probing_2024}, from which the quantum weight can be determined by the optical sum rule. Comparing these experimental results and theoretical calculations will provide a nontrivial test of our theory.}

Importantly, the concept of quantum weight also applies to strongly correlated systems, including Mott insulators~\cite{aebischer_dielectric_2001} and fractional Chern insulators, which cannot be described by band theory. Very recently, the quantum weights of various fractional Chern insulators in twisted MoTe$_2$ have been calculated and compared with the corresponding topological bounds~\cite{zaklama_structure_2024}. These results will provide valuable insights into optical excitations in this system.  

Another exciting direction is to study the effect of disorder on the quantum weight, especially near a topological phase transition. Numerical study of this problem has just started~\cite{alisultanov_disorder-induced_2024}. 

We also highlight that the study of static structure factor and quantum weight has stimulated further research on quantum geometry,  which plays important roles in linear~\cite{jankowski_optical_2024, mao_is_2024} and nonlinear optical responses~\cite{Ahn2020, ahn_riemannian_2022, bradlyn_spectral_2024, postlewaite_nonlinear_2024}, as well as excitonic effects~\cite{qiu_quantum_2024}. 

It is tantalizing to explore possible connections between quantum weight and quantum entanglement in periodic systems, both of which are quantum in nature. In this regard, we note that quantum entanglement and particle number fluctuation are often intertwined~\cite{klich_quantum_2009, song_entanglement_2011, levine_detecting_2011, cramer_measuring_2011, song_bipartite_2012, tam_topological_2022}. Recent studies suggest that in two-dimensional insulators, the corner contribution to particle number fluctuation is closely related to both entanglement entropy and quantum weight~\cite{estienne_cornering_2022, tam_corner_2024, wu_corner_2024, kruchkov_entanglement_2024}, hinting at a deeper connection between the two. 
Since the quantum weight can be experimentally determined through the static structure factor, exploring its relationship with quantum entanglement may offer a direct way to measure the latter.  

Lastly, we note that the concept of quantum weight naturally generalizes to general $U(1)$ symmetric systems. For example, when the total spin along $z$-direction is conserved, the system has spin $U(1)$ symmetry and the spin density $s^z(\vec{r})$ can be regarded as $U(1)$ charge density. Accordingly, $q^2$ coefficient of the spin structure factor $S^s_{\vec{q}}=(1/V)\expval{s^z_{\vec{q}}s^z_{-\vec{q}}}$ defines the spin analog of quantum weight, which may be called ``spin quantum weight''. Similarly to the topological bound on quantum weight, one can show the lower bound on the spin quantum weight as well~\cite{onishi_topological_2024}. Spin quantum weight would be useful for studying superconductors and $U(1)$ chiral spin liquids.

\begin{acknowledgements}
We thank Ivo Souza, Richard Martin and especially Raffaele Resta for valuable communications. We thank Ling Miao for her suggestion on the title of this paper.  
This work was  supported by National Science Foundation (NSF) Convergence Accelerator Award No. 2235945 and a Simons Investigator Award from the Simons Foundation. YO is grateful for the support provided by the Funai Overseas Scholarship. LF was supported in part by  the Air Force Office of Scientific Research under award number FA2386-24-1-4043.     
\end{acknowledgements}

\bibliography{references_zotero}

\appendix
\begin{widetext}
\section*{Supplemental Material}
\section{Calculation for 3D Coulomb solids}
\subsection{Structure factor for strongly localized electrons}
Here, we calculate the structure factor for the 3D Coulomb solids, i.e., the strongly localized electrons described in the main text. The effective Lagrangian is given by (also see the supplemental materials in Ref.~\cite{onishi_universal_2024})
\begin{align}
    {\cal L} &= \sum_{\vec q} \frac{m}{2}\qty(\abs{\dot{\vec{u}}_{\vec{q}}}^2 - (\omega_0^2 + \omega_p^2 \frac{q_\alpha q_\beta}{q^2}) u_{\vec{q}\alpha}u_{-\vec{q}\beta}), \label{eq:Ham}
\end{align}
where  $\vec{u}_{\vec{q}} = (1/\sqrt{N})\sum_{\vec{q}}e^{-i\vec{q}\vdot\vec{R}_i}u_{i}$ is the Fourier transform of the displacement of the $i$th electron $\vec{u}_i$ from the equilibrium position $\vec{R}_i$, with $N$ the total number of unit cells. 

The number density operator $n(\vec{r})$ is given by 
    $n(\vec{r}) = \sum_{i} \delta(\vec{r}-(\vec{R}_i + \vec{u}_i))$
and thus its Fourier transform $n_{\vec{q}}$ is given by 
\begin{align}
    n_{\vec{q}} &= \int\dd{\vec{r}} e^{-i\vec{q}\vdot\vec{r}} n(\vec{r}) = \sum_{i} e^{-i\vec{q}\vdot(\vec{R}_i+\vec{u}_i)}
\end{align}
Therefore, for finite $\vec{q}$ with the expectation value $\expval{n_{\vec{q}}}=0$, the static structure factor is given by 
\begin{align}
    S_{\vec{q}}&\equiv\frac{1}{V}\expval{n_{\vec{q}}n_{-\vec{q}}} = \frac{1}{V}\sum_{i,j}e^{-i\vec{q}\vdot(\vec{R}_i-\vec{R}_j)}\expval{e^{-i\vec{q}\vdot\vec{u}_i} e^{i\vec{q}\vdot\vec{u}_j}} \simeq \frac{1}{V}\sum_{i,j}e^{-i\vec{q}\vdot(\vec{R}_i-\vec{R}_j)}\expval{(\vec{q}\vdot\vec{u}_i)(\vec{q}\vdot\vec{u}_j)} + \order{q^3}\\
    &=  n \expval{(\vec{q}\vdot\vec{u}_{\vec{q}})(\vec{q}\vdot\vec{u}_{-\vec{q}})}+ \order{q^3}.
\end{align}
This is the expression in the main text.

\subsection{Response functions}
Here we calculate the response functions of 3D Coulomb solids, described by the following Lagrangian: 
\begin{align}
    {\cal L} &= \sum_{\vec q} \frac{m}{2}\qty(\abs{\dot{\vec{u}}_{\vec{q}}}^2 - (\omega_0^2 + \omega_p^2 \frac{q_\alpha q_\beta}{q^2}) u_{\vec{q}\alpha}u_{-\vec{q}\beta}), \label{ap:eq:Lag}
\end{align}
where $\vec{u}_{\vec{q}}=(1/\sqrt{N})\sum_i e^{-i\vec{q}\vdot\vec{R}_i} \vec{u}_i$, and $\vec{u}_i$ is the displacement of the electron in $i$-th unit cell from potential minima $\vec{R}_i$. $N$ is the total number of unit cells in the system. The corresponding Hamiltonian is given by 
\begin{align}
	H &= \sum_{\vec{q}} \frac{\abs{\vec{P}_{\vec{q}}}^2}{2m} + \frac{m}{2}(\omega_0^2 + \omega_p^2 \frac{q_\alpha q_\beta}{q^2}) u_{\vec{q}\alpha}u_{-\vec{q}\beta} \nonumber \\
	&= \sum_{\vec{q}} \qty[\frac{\abs{P^L_{\vec{q}}}^2}{2m} + \frac{m\omega_L^2}{2}\abs{u^L_{\vec{q}}}^2 + \frac{\abs{P^T_{\vec{q}}}^2}{2m} + \frac{m\omega_T^2}{2}\abs{\vec{u}^T_{\vec{q}}}^2] \label{ap:eq:Ham}
\end{align}
where $\vec{P}_{\vec{q}}=\partial {\cal L}/\partial \dot{\vec{u}_{\vec{q}}} = m\dot{\vec{u}}_{-\vec{q}}$ is the canonical momentum conjugate to $\dot{\vec{u}}_{\vec{q}}$. $\vec{u}_{\vec{q}}^L$ and $\vec{u}_{\vec{q}}^T$ are the longitudinal and transverse components of $\vec{u}_{\vec{q}}$ with respect to $\vec{q}$, respectively, and similar for $P^L_{\vec{q}}$, $P^T_{\vec{q}}$. The frequencies are given by $\omega_L=\sqrt{\omega_0^2+\omega_p^2}$ and $\omega_T=\omega_0$. 
Our interest here is in the response to an electric field $\vec{E}$, either longitudinal ($\vec{E}_{\vec{q}}^L\parallel \vec{q}$) or transverse ($\vec{E}_{\vec{q}}^T\perp\vec{q}$), with the Fourier transfrom $\vec{E}_{\vec{q}}=\int\dd^3{\vec{r}}e^{-i\vec{q}\vdot\vec{r}}\vec{E}(\vec{r})$. The electric field is coupled to the displacement $\vec{u}_i$ as 
\begin{align}
    -e\sum_i \vec{u}_i\vdot\vec{E}(\vec{R}_i) = - \frac{ne}{\sqrt{N}} \sum_{\vec{q}} \vec{u}_{-\vec{q}}\vdot\vec{E}_{\vec{q}} = - \frac{ne}{\sqrt{N}} \sum_{\vec{q}} \qty(\vec{u}^L_{-\vec{q}}\vdot\vec{E}^L_{\vec{q}}+\vec{u}^T_{-\vec{q}}\vdot\vec{E}^T_{\vec{q}}). \label{ap:eq:couple_to_ext}
\end{align}
Eq.~\eqref{ap:eq:couple_to_ext} explicitly shows that the longitudinal (transverse) field couples to the longitudinal (transverse) mode. 

Since this system is described by the harmonic oscillator for each mode $\vec{u}^{L}_{\vec{q}}, \vec{u}^{T}_{\vec{q}}$ with frequency $\omega_L=\sqrt{\omega_0^2+\omega_p^2}, \omega_T=\omega_0$, we can easily calculate the response of this function. It is convenient to consider the response of $\vec{u}^{L/T}_{\vec{q}}$ to an electric field $\vec{E}^{L/T}_{\vec{q}}$ at $\vec{q}\to 0$. Defining the response function $\tilde{\chi}^L, \tilde{\chi}^T$ as $\vec{u}^{L/T}_{\vec{q}}(\omega) = \tilde{\chi}^{L/T}_{\vec{q}}(\omega)\vec{E}^{L/T}_{\vec{q}}(\omega)$ with the fourier transform defined as $\vec{E}_{\vec{q}}(\omega) = \int\dd{t}e^{i\omega t}\vec{E}_{\vec{q}}(t)$, we find 
\begin{align}
	\tilde{\chi}^{L/T}_{\vec{q}}(\omega) &= \frac{1}{\sqrt{N}}\frac{ne}{m(\omega_{L/T}^2-\omega_+^2)} \quad (\vec{q}\to 0) \label{ap:eq:chiLT}
\end{align}
where $\omega_+=\omega+i\eta$ with $\eta\to 0^+$.
The factor $1/\sqrt{N}$ comes from the normalization of the displacement field.

Noting that the density is given by $\rho_{\vec{q}}= e\sqrt{N}(-i\vec{q}\vdot\vec{u}_{\vec{q}})+\order{q^2}$ at $\vec{q}$ away from the reciprocal lattice vector, we find the density response function $\Pi(\vec{q}, \omega)$ at the leading order in $q$ is given by $\Pi(\vec{q}\to 0,\omega) = -eq^2\sqrt{N}\tilde{\chi}_{\vec{q}}^L(\omega)$ and thus 
\begin{align}
    \Pi(\vec{q}\to 0, \omega) = - q^2\frac{ne^2}{m(\omega_L^2-\omega^2)} = -\epsilon_0 q^2 \frac{\omega_p^2}{\omega_L^2-\omega_+^2}.
\end{align}
Note that the sum over $\vec{q}$ in Eq.~\eqref{ap:eq:Lag} runs over both $\vec{q}$ and $-\vec{q}$ to get the correct numerical factor. Since the potential generates only the longitudinal field $\vec{E}^L$, $\Pi$ is governed by the response of the longitudinal mode $\vec{u}^L$, resulting in the pole at $\omega_L$. From $\Pi$, we can also calculate the dielectric function $\epsilon(\vec{q},\omega)$ as 
\begin{align}
    \epsilon(\vec{q}, \omega) = (1+U(\vec{q})\Pi(\vec{q},\omega))^{-1} \to \frac{\omega_L^2-\omega_+^2}{\omega_T^2-\omega_+^2} \quad (\vec{q}\to 0) \label{ap:eq:eps}
\end{align}

We can also calculate the optical conductivity. The Fourier transform of the current operator $\vec{j}(\vec{r})=e\sum_i \dot{\vec{u}}_i\delta(\vec{r}-\vec{r}_i)$ is given by $\vec{j}_{\vec{q}}=\int\dd^3{\vec{r}}e^{-i\vec{q}\vdot\vec{r}}\vec{j}(\vec{r}) = e\sqrt{N}\dot{\vec{u}}_{\vec{q}}$. Noting that the optical field is transverse, the optical conductivity $\sigma$ is the response of the current to the transverse field: $\vec{j}(\vec{q}, \omega)=\sigma(\vec{q},\omega)\vec{E}^T_{\vec{q}}(\omega)$, we find $\sigma(\vec{q},\omega) = -ie\sqrt{N}\omega \tilde{\chi}^T_{\vec{q}}(\omega)$ and in $\vec{q}\to0$ limit it is given by 
\begin{align}
	\sigma(\omega) = -i\omega \frac{ne^2}{m(\omega_T^2-\omega_+^2)} = -i\omega \epsilon_0 \frac{\omega_p^2}{\omega_T^2-\omega_+^2} \label{ap:eq:sigma_chiT}
\end{align}
One can also obtain the optical conductivity from the longitudinal response at $\vec{q}\to 0$ by using Eq.~\eqref{ap:eq:eps}:
\begin{align}
    \sigma^L_{\alpha\beta}(\vec{q},\omega) & = \frac{-i\omega}{U(\vec{q})q_{\alpha}q_{\beta}} (\epsilon(\vec{q},\omega)-1) \to -i\omega \epsilon_0\frac{\omega_p^2}{\omega_T^2-\omega_+^2} \label{ap:eq:sigma_epsilon_new}
\end{align}
Both expressions~\eqref{ap:eq:sigma_chiT}, \eqref{ap:eq:sigma_epsilon_new} are consistent.

\section{Density and optical responses in Coulomb systems}
As discussed in the main text, the relation between the density and optical responses in Coulomb systems is subtle compared to systems with short-range interactions. Although these are discussed in detail in textbooks (for example, see Ref.~\cite{pines2018theory}), we describe some of them here for the sake of completeness.

The density response function $\Pi$ defines the dielectric function $\epsilon(\omega)$ which relates the \textit{external} charge density $\rho_{\rm ext}$ to the \textit{total} (or screened) charge density $\rho_{\rm tot}$
as $\rho_{\rm ext} = \epsilon \rho_{\rm tot}$ (for the moment we assume the dielectric tensor is isotropic and present the general result later). The total charge is the sum of the external charge and the induced charge: $\rho_{\rm tot} = \rho_{\rm ext} + \Pi(\vec{q})V(\vec{q},\omega)$, where $V(\vec{q},\omega)=U(\vec{q})\rho_{\rm ext}(\vec{q},\omega)$ is the potential created by external charges, and $U(\vec{q})$ describes the interaction between two point charges. This leads to a relation between $\epsilon$ and $\Pi$: 
\begin{align}
    \epsilon(\vec{q},\omega) &= (1+U(\vec{q})\Pi(\vec{q},\omega))^{-1} \label{eq:eps-Pi_new}
\end{align}
Typically, $\Pi(\vec{q},\omega)\propto q^2$ at $\vec{q}\to 0$ in the gapped systems, and thus the short-range interaction satisfying $U(\vec{q})q^2 \to 0$ at $\vec{q}\to 0$ results in $\epsilon(\omega)=1$. However, for 3D Coulomb interaction $U(\vec{q})=1/(\epsilon_0 q^2)$, $\epsilon(\omega)\neq 1$ even at $\vec{q}\to 0$, reflecting the screening effect due to the long-range nature of Coulomb interaction. 

$\sigma^L$ can be expressed with $\epsilon(\vec{q},\omega)$ as well. As discussed in the main text, $\sigma^L$ is related to the density response function $\Pi$ as 
\begin{align}
    \Pi(\vec{q},\omega) &= -i\frac{q_{\alpha}q_{\beta}\sigma^L_{\alpha\beta}(\vec{q},\omega)}{\omega\epsilon(\vec{q},\omega)} \label{ap:eq:Pi_sigma_new}. 
\end{align}
With Eq.~\eqref{eq:eps-Pi_new} and Eq.~\eqref{ap:eq:Pi_sigma_new}, we can relate $\sigma^L$ with the dielectric function $\epsilon$ as 
\begin{align}
    \sigma^L_{\alpha\beta}(\vec{q},\omega) & = \frac{-i\omega}{U(\vec{q})q_{\alpha}q_{\beta}} (\epsilon(\vec{q},\omega)-1) \nonumber \\
    &\to -i\omega\epsilon_0(\epsilon(\omega)-1), \quad (\vec{q}\to 0) \label{eq:sigma_epsilon_new}
\end{align}
where we have used $U(\vec{q})=1/(\epsilon_0q^2)$ in the second line. Since $\sigma^L(\vec{q}, \omega)$ reduces to the optical conductivity at $\vec{q}=0$, the last expression is the common expression relating the dielectric constant or the polarizability to the optical conductivity.

\section{Structure factor for noninteracting band insulators}
In this section, we calculate the static structure factor for general noninteracting band insulators. Writing the cell-periodic Bloch wavefunction for $n$th band as $\ket{u_{n\vec{k}}}$ with the wavevector $\vec{k}$, the number density operator with wavevector $\vec{q}$ is given by 
\begin{align}
    n_{\vec{q}} &= \sum_{n, m} \sum_{\vec{k}} \ip{u_{n,\vec{k}}}{u_{m,\vec{k}+\vec{q}}} \cre{c}_{n{\vec{k}}}\ani{c}_{m,\vec{k}+\vec{q}},
\end{align}
with $\ani{c}_{n,\vec{k}}$ the annihilation (creation) operator of electron in $n$th band with wavevector $\vec{k}$. Then the static structure factor for $\vec{q}$ away from reciprocal lattice vectors is given by 
\begin{align}
    S_{\vec{q}}&\equiv\frac{1}{V}\expval{n_{\vec{q}}n_{-\vec{q}}} = \frac{1}{V}\sum_{n,m,\vec{k}}\sum_{n',m',\vec{k}'}\ip{u_{n,\vec{k}}}{u_{m,\vec{k}+\vec{q}}}\ip{u_{n',\vec{k}'}}{u_{m',\vec{k}'+\vec{q}'}}\expval{\cre{c}_{n{\vec{k}}}\ani{c}_{m,\vec{k}+\vec{q}}  \cre{c}_{n'{\vec{k}'}}\ani{c}_{m',\vec{k}'+\vec{q}'}} \\
    &= \frac{1}{V}\sum_{n,m,\vec{k}}\ip{u_{n,\vec{k}}}{u_{m,\vec{k}+\vec{q}}}\ip{u_{m,\vec{k}+\vec{q}}}{u_{n,\vec{k}}}(1-f_m(\vec{k}+\vec{q}))f_n(\vec{k})
\end{align}
where $f_n(\vec{k}) = \expval{\cre{c}_{n\vec{k}}\ani{c}_{n\vec{k}}}$ is the occupation number of $n$th band at wavevector $\vec{k}$. Defining the projection operator $P(\vec{k}) = \sum_{n}^{\rm occ} \ket{u_{n\vec{k}}}f_n(\vec{k})\bra{u_{n\vec{k}}}$, we obtain an expression of $S_{\vec{q}}$ at $\vec{q}$ away from the reciprocal lattice vectors for a general noninteracting system as 
\begin{align}
    S_{\vec{q}}&= \frac{1}{V}\sum_{\vec{k}}\Tr[P(\vec{k})(1-P(\vec{k}+\vec{q}))] = \frac{1}{V}\sum_{\vec{k}}\Tr[P(\vec{k})(P(\vec{k})-P(\vec{k}+\vec{q}))]
\end{align}
where we have used $P(\vec{k})^2 = P(\vec{k})$ in the last equality. Further rewriting the summation over $\vec{k}$ with integral, we obtain:
\begin{align}
    S_{\vec{q}}&= \int_{\rm BZ}\frac{\dd^d\vec{k}}{(2\pi)^d}\Tr[P(\vec{k})(P(\vec{k})-P(\vec{k}+\vec{q}))]. \label{eq:Sq_BI_P}
\end{align}
In particular, for band insulators, the projection operator becomes $P(\vec{k})=\sum_{n}^{\rm occ} \ket{u_{n\vec{k}}}\bra{u_{n\vec{k}}}$ and Eq.~\eqref{eq:Sq_BI_P} reduces to the expression in the main text.

\section{Formula for structure factor in real space}
Sometimes, it is easy to calculate the structure factor in real space. Here, we provide the formula of the structure factor of noninteracting systems in terms of real space, rather than $k$-space. 

The equal-time density-density correlation function $S(\vec{r}_1, \vec{r}_2)\equiv \expval{\rho(\vec{r}_1)\rho(\vec{r}_2)}-\expval{\rho(\vec{r}_1)}\expval{\rho(\vec{r}_2)}$ is given by 
\begin{align}
    S(\vec{r}_1, \vec{r}_2) &= (\delta(\vec{r}_1,\vec{r}_2)-P(\vec{r}_1, \vec{r}_2)) P(\vec{r}_2, \vec{r}_1). \label{eq:Sq_P}
\end{align}
$P(\vec{r}_1, \vec{r}_2)=\mel{\vec{r}_1}{P}{\vec{r}_2}$ is the matrix element of the projection operator onto the occupied states, given by 
\begin{align}
    P &= \sum_{a:{\rm occ}}\ketbra{\psi_a}{\psi_a},  \\
    P(\vec{r}_1, \vec{r}_2) &= \sum_{a:{\rm occ}}\psi_a(\vec{r}_1)\psi_a^*(\vec{r}_2)
\end{align}
where $\psi_a$ is a normalized occupied state orthogonal to each other, i.e., $\ip{\psi_a}{\psi_b}=\delta_{ab}$. If one chooses the Bloch wavefunctions as $\psi_a$, one obtains   
\begin{align}
    P &= \sum_{n,\vec{k}}\ket{\phi_{n\vec{k}}} f_n(\vec{k}) \bra{\phi_{n\vec{k}}} \\
    P(\vec{r}_1, \vec{r}_2) &= \sum_{n,\vec{k}}\phi_{n\vec{k}}(\vec{r}_1) f_n(\vec{k}) \phi_{n\vec{k}}^*(\vec{r}_2)
\end{align}
where $\ket{\phi_{n\vec{k}}}$ is the Bloch wavefunction with $\phi_{n\vec{k}}(\vec{r})=\ip{\vec{r}}{\phi_{n\vec{k}}}$. Note that $\phi_{n\vec{k}}(\vec{r}+\vec{a})=e^{i\vec{k}\vdot\vec{a}}\phi_{n\vec{k}}(\vec{r})$ and $\phi_{n\vec{k}}$ is \textit{not} the cell-periodic wavefunction $u_{n\vec{k}}=e^{-i\vec{k}\vdot\vec{r}}\phi_{n\vec{k}}(\vec{r})$. $\phi_{n\vec{k}}$ is normalized with respect to integral over the entire system $V$: $\int_V\dd{\vec{r}}\phi^{*}_{n\vec{k}_1}(\vec{r})\phi_{m\vec{k}_2}(\vec{r}) = \delta_{nm}\delta_{\vec{k}_1,\vec{k}_2}$.

From the density-density correlation function in real space $S(\vec{r}_1, \vec{r}_2)$, we can obtain its Fourier transform $S'_{\vec{q}}$ as 
\begin{align}
    {S'}_{\vec{q}} &= \frac{1}{V}\int\dd{\vec{r}_1}\dd{\vec{r}_2} e^{-i\vec{q}\vdot(\vec{r}_1-\vec{r}_2)} S(\vec{r}_1, \vec{r}_2) \label{eq:Sq_Sr}
\end{align}
where $S'_{\vec{q}}=(1/V)(\expval{\rho_{\vec{q}}\rho_{-\vec{q}}}-\expval{\rho_{\vec{q}}}\expval{\rho_{-\vec{q}}})$. $S'_{\vec{q}}$ differs from the static structure factor $S_{\vec{q}}=(1/V)\expval{\rho_{\vec{q}}\rho_{-\vec{q}}}$ when $\vec{q}$ is a reciprocal lattice vector, but is identical with $S_{\vec{q}}$ at generic $\vec{q}$.

\section{Structure factor of an array of harmonic oscillators}
Consider an array of $d$-dimensional harmonic oscillators with potential $V(\vec{r})=m\omega_0^2 r^2/(2m)$, neglecting the hopping among them. When each harmonic oscillator contains one electron, the ground state of this system is given by the slater determinant of the wavefunctions for each oscillator at $\vec{R}$: $\psi(\vec{r}-\vec{R}) =  \qty(m\omega_0/(2\pi\hbar))^{1/4} \exp(-m\omega_0 \abs{\vec{r}-\vec{R}}^2/(2\hbar))$. 

To calculate the structure factor, we first calculate the projector $P(\vec{r}_1, \vec{r}_2)$. $P(\vec{r}_1, \vec{r}_2)$ in this case is given by 
\begin{align}
    P(\vec{r}_1,\vec{r}_2) &= \qty(\frac{m\omega_0}{2\pi\hbar})^{d/2}\sum_{\vec{R}}\exp[-\frac{m\omega_0}{2\hbar}\qty(\abs{\vec{r}_1-\vec{R}}^2+\abs{\vec{r}_2-\vec{R}}^2)]
\end{align}

In this case, the structure factor is given by Eq.~\eqref{eq:Sq_Sr} and \eqref{eq:Sq_P} as
\begin{align}
    S_{\vec{q}} = n(1-e^{-\frac{\hbar q^2}{2m\omega_0}}), 
\end{align}
where $n=N/V$ is the electron density with $N$ the number of unit cells and $V$ the volume of the system. Noting that each oscillator contains one electron and thus $n=1/V_{UC}$, this reduces to the expression in the main text.

\section{Table for the experimental parameters}
\renewcommand{\cellalign}{cl}
\renewcommand{\arraystretch}{1.7}
\begin{table}[h!]
\centering
\begin{tabular}{l|cccccccc|l}
Material & $\epsilon(\infty)$ & $n$ [$10^{30}$\si{m^{-3}}] & $E_g^T$ [\si{\electronvolt}] & $E_g^{T,m}$ [\si{\electronvolt}] & $E_g^T/E_g^{T,m}$ & $E_g^L$ [\si{\electronvolt}] & $E_g^{L,m}$ [\si{\electronvolt}] & $E_g^L/E_g^{L,m}$ & Comment\\
\hline
c-BN & 4.46~\cite{eremets_optical_1995} & \SI{1.02}{}~\cite{madelung_semiconductors_2004}  & 14.5~\cite{madelung_semiconductors_2004}  & 20.11 & 0.721 & 30.4~\cite{jaouen_eels_1995} & 42.48 & 0.716 & $E_g^T$ is direct gap.\\
Si & 12.0~\cite{madelung_semiconductors_2004} & \SI{0.700}{}~\cite{madelung_semiconductors_2004} & 4.18~\cite{madelung_semiconductors_2004} & 9.38 & 0.446 & 16.7~\cite{egerton_electron_2008} & 32.44 & 0.515 & $E_g^T$ is direct gap.\\
Diamond & 5.7~\cite{madelung_semiconductors_2004} & \SI{1.06}{}~\cite{madelung_semiconductors_2004} & 7.1~\cite{logothetidis_origin_1992}  & 17.62 & 0.403 & 34.0~\cite{bursill_plasmon_1997} & 42.06 & 0.808 & $E_g^T$ is direct gap.\\
LiCl & 2.78~\cite{Ashcroft1976} & \SI{0.589}{}\cite{bendow_pressure_1974} & 8.6~\cite{teegarden_optical_1967} & 21.36 & 0.403 & - & 35.62 & - & $E_g^T$ is from the lowest absorption peak \\
LiF & 1.96~\cite{Ashcroft1976} & \SI{0.739}{}\cite{bendow_pressure_1974} & 12.6~\cite{roessler_electronic_1967} & 32.58 & 0.387 & - & 45.61 & - & $E_g^T$ is an excitonic gap\\
3C-SiC & 6.38~\cite{madelung_semiconductors_2004} & \SI{0.965}{}~\cite{madelung_semiconductors_2004} & 6.0~\cite{madelung_semiconductors_2004} & 15.73 & 0.381 & 21.8~\cite{costantini_analysis_2023} & 39.73 & 0.549 & $E_g^T$ is direct gap.\\
AlN & 4.93~\cite{madelung_semiconductors_2004} & \SI{0.958}{}~\cite{madelung_semiconductors_2004} & 6.19~\cite{madelung_semiconductors_2004} & 18.34 & 0.338 & - & 40.72 & - & $E_g^T$ is direct gap.. $\epsilon(\infty)=\epsilon_{\parallel}(\infty)$ \\
NaCl & 2.34~\cite{Ashcroft1976} & \SI{0.624}{}~\cite{rumble_crc_2023} & 7.9~\cite{teegarden_optical_1967} & 25.34 & 0.312 & 15.5~\cite{egerton_electron_2008} & 38.77 & 0.400 & $E_g^T$ is from the lowest absorption peak\\
KCl & 2.19~\cite{Ashcroft1976} & \SI{0.578}{}~\cite{rumble_crc_2023} & 7.8~\cite{teegarden_optical_1967} & 25.87 & 0.301 & 14.1~\cite{akkerman_inelastic_1996} & 38.29 & 0.368 & \makecell{$E_g^T$ is from the lowest absorption peak. \\ $E^L_g$ is the first prominent peak in EELS.} \\
LiBr & 3.17~\cite{Ashcroft1976} & \SI{0.914}{}\cite{bendow_pressure_1974} & 7.2~\cite{teegarden_optical_1967} & 24.09 & 0.299 & - & 42.90 & - & $E_g^T$ is from the lowest absorption peak\\
KF & 1.85~\cite{Ashcroft1976} & \SI{0.736}{}\cite{bendow_pressure_1974} & 9.8~\cite{teegarden_optical_1967} & 34.54 & 0.284 & - & 46.98 & - & $E_g^T$ is from the lowest absorption peak\\
MgO & 2.94~\cite{madelung_semiconductors_2004} & \SI{1.07}{}~\cite{madelung_semiconductors_2004} & 7.67~\cite{madelung_semiconductors_2004} & 27.52 & 0.279 & 22.3~\cite{egerton_electron_2008} & 47.21 & 0.472 & $E_g^T$ is an excitonic gap\\
NaF & 1.74~\cite{Ashcroft1976} & \SI{0.804}{}~\cite{rumble_crc_2023} & 10.6~\cite{teegarden_optical_1967} & 38.70 & 0.274 & - & 51.05 & - & $E_g^T$ is from the lowest absorption peak\\
RbF & 1.96~\cite{Ashcroft1976} & \SI{1.03}{}\cite{bendow_pressure_1974} & 9.5~\cite{teegarden_optical_1967} & 38.38 & 0.248 & - & 53.73 & - & $E_g^T$ is from the lowest absorption peak \\
NaBr & 2.59~\cite{Ashcroft1976} & \SI{0.869}{}\cite{bendow_pressure_1974} & 6.7~\cite{teegarden_optical_1967} & 27.45 & 0.244 & - & 44.18 & - & $E_g^T$ is from the lowest absorption peak\\
KBr & 2.34~\cite{Ashcroft1976} & \SI{0.758}{}\cite{bendow_pressure_1974} & 6.7~\cite{teegarden_optical_1967} & 27.93 & 0.240 & 13.2~\cite{akkerman_inelastic_1996} & 42.73 & 0.309 & \makecell{$E_g^T$ is from the lowest absorption peak. \\ $E_g^L$ is the first prominent peak in EELS.} \\
KI & 2.62~\cite{Ashcroft1976} & \SI{0.818}{}\cite{bendow_pressure_1974} & 5.8~\cite{teegarden_optical_1967} & 26.39 & 0.220 & 11.8~\cite{akkerman_inelastic_1996}& 42.72 & 0.276 & \makecell{$E_g^T$ is from the lowest absorption peak. \\ $E_g^L$ is the first prominent peak in EELS.} \\
RbBr & 2.34~\cite{Ashcroft1976} & \SI{0.892}{}\cite{bendow_pressure_1974} & 6.6~\cite{teegarden_optical_1967} & 30.30 & 0.218 & - & 46.35 & - & $E_g^T$ is from the lowest absorption peak\\
NaI & 2.93~\cite{Ashcroft1976} & \SI{0.950}{}\cite{bendow_pressure_1974} & 5.6~\cite{teegarden_optical_1967} & 26.05 & 0.215 & - & 44.58 & - & $E_g^T$ is from the lowest absorption peak\\
RbI & 2.59~\cite{Ashcroft1976} & \SI{0.918}{}\cite{bendow_pressure_1974} & 5.7~\cite{teegarden_optical_1967} & 28.21 & 0.202 & - & 45.40 & - & $E_g^T$ is from the lowest absorption peak\\
$\alpha$-GaN & 5.2~\cite{madelung_semiconductors_2004} & \SI{1.66}{}~\cite{madelung_semiconductors_2004} & 3.48~\cite{madelung_semiconductors_2004} & 23.36 & 0.149 & - & 53.27 & - &  $E_g^T$ is A-exciton. $\epsilon(\infty)=\epsilon_\perp(\infty).$\\
ZnO & 3.75~\cite{madelung_semiconductors_2004} & \SI{1.60}{}~\cite{madelung_semiconductors_2004} & 3.44~\cite{madelung_semiconductors_2004} & 28.30 & 0.122 & 18.2~\cite{huang_characterization_2011} & 54.81 & 0.332 & $\epsilon(\infty)=\epsilon_\parallel(\infty).$\\
GaAs & 10.9~\cite{madelung_semiconductors_2004} & \SI{1.42}{}~\cite{madelung_semiconductors_2004} & 1.52~\cite{madelung_semiconductors_2004} & 14.08 & 0.108 & 15.8~\cite{horak_cerenkov_2015} & 46.38 & 0.341 & $E_g^T$ is direct gap.\\
InN & 8.4~\cite{madelung_semiconductors_2004} & \SI{1.80}{}~\cite{madelung_semiconductors_2004} & 1.95~\cite{madelung_semiconductors_2004} & 18.34 & 0.106 & - & 53.15 & - & $E_g^T$ is direct gap.
\\
Ge & 16.0~\cite{madelung_semiconductors_2004} & \SI{1.41}{}~\cite{madelung_semiconductors_2004} & 0.898~\cite{madelung_semiconductors_2004} & 11.40 & 0.079 & 16.0~\cite{poursoti_deep_2022} & 45.59 & 0.351 & $E_g^T$ is direct gap.\\
SnTe & 40 \cite{suzuki_optical_1995,schoolar_optical_1968} & \SI{1.61}{}~\cite{madelung_semiconductors_2004} & 0.36~\cite{madelung_semiconductors_2004} & 7.55 & 0.048 & - & 47.75 & - & \makecell{$E_g^T$ is direct gap.  $\epsilon(\infty)$ varies \\ among Ref.~\cite{suzuki_optical_1995, schoolar_optical_1968}  around 40-50.} \\
Sb$_2$Te$_3$ & 51.0~\cite{madelung_semiconductors_2004} & \SI{1.69}{}~\cite{madelung_semiconductors_2004} & 0.28~\cite{madelung_semiconductors_2004} & 6.83 & 0.041 & - & 48.75 & - & $E_g^T$ is direct gap. $\epsilon(\infty)=\epsilon_\perp(\infty).$\\
Bi$_2$Te$_3$ & 85.0~\cite{madelung_semiconductors_2004} & \SI{1.90}{}~\cite{madelung_semiconductors_2004} & 0.21~\cite{nemov_band_2019} & 5.58 & 0.038 & - & 51.49 & - & $\epsilon(\infty)=\epsilon_\perp(\infty).$\\
InAs & 12.4~\cite{madelung_semiconductors_2004} & \SI{1.48}{}~\cite{madelung_semiconductors_2004} & 0.418~\cite{madelung_semiconductors_2004}& 13.37 & 0.031 & 13.9~\cite{kundmann_study_1988} & 47.04 & 0.295 &   $E_g^T$ is direct gap.\\
Bi$_2$Se$_3$ & 29.0~\cite{madelung_semiconductors_2004} & \SI{1.89}{}~\cite{madelung_semiconductors_2004} & 0.22~\cite{martinez_determination_2017} & 9.65 & 0.023 & - & 51.98 & - & $\epsilon(\infty)=\epsilon_\perp(\infty).$ \\
Li & - & \SI{0.139}{}~\cite{rumble_crc_2023} & - & - & - & 7.0~\cite{mauchamp_local_2008} & 13.84 & 0.506 & Metal \\
Be & - & \SI{0.493}{}~\cite{rumble_crc_2023} & - & - & - & 18.7~\cite{egerton_electron_2011} & 26.08 & 0.717 & Metal \\
Na & - & \SI{0.276}{}~\cite{rumble_crc_2023} & - & - & - & 19.6~\cite{egerton_electron_2011} & 26.08 & 0.291 & Metal \\
Mg & - & \SI{0.516}{}~\cite{rumble_crc_2023} & - & - & - & 10.3~\cite{egerton_electron_2011} & 26.68 & 0.386 & Metal \\

\end{tabular}
\caption{Parameters used to calculate the gap bound in Fig.~\ref{fig:Kbound_CS}. $\epsilon(\infty)$ is the optical dielectric constant, $n$ is the electron density, $E_g^T$ is the optical gap, $E_g^L$ is the measured plasma frequency in electron energy loss spectroscopy (EELS), and $E_g^{T/L,m}$ is the gap bound calculated following Ref.~\cite{onishi_universal_2024}.}
\label{tab:parameters}
\end{table}

\end{widetext}

\end{document}